\documentclass[aps,prl,superscriptaddress,floats,twocolumn,nofootinbib]{revtex4}
\usepackage{graphicx}% Include figure files
\usepackage{dcolumn}% Align table columns on decimal point
\usepackage{bm}% bold math
\usepackage[english]{babel}
\usepackage[latin1]{inputenc}
\usepackage{graphicx}
\usepackage{epstopdf}
\usepackage{amsfonts}
\usepackage{comment}
\usepackage{url}
\usepackage{hyperref}
\usepackage{color}
\usepackage{epsfig}
\usepackage{mathrsfs}
\usepackage{amsmath,amssymb}
\usepackage{subfigure}

\usepackage{framed}
\usepackage[svgnames]{xcolor}
\definecolor{shadecolor}{named}{LightGrey}

\newcommand{\lto}[1]{\longrightarrow#1}

\renewcommand{\(}{\left(}
\renewcommand{\)}{\right)}
\renewcommand{\[}{\left[}
\renewcommand{\]}{\right]}

\newcommand{\Eq}[1]{Eq.\,(\ref{#1})}
\newcommand{\Fig}[1]{Fig.\,\ref{#1}}

\begin{document}

\graphicspath{{figure/}}
\selectlanguage{english}

\title{Random Walks on Multiplex Networks: Supplementary Information for ``Navigability of Interconnected Networks under Random Failures''}

\author{Manlio De Domenico}
\affiliation{Departament d'Enginyeria Inform\'atica i Matem\'atiques, Universitat Rovira I Virgili, Tarragona, Spain}

\author{Albert Sol\'e-Ribalta}
\affiliation{Departament d'Enginyeria Inform\'atica i Matem\'atiques, Universitat Rovira I Virgili, Tarragona, Spain}

\author{Sergio G\'omez}
\affiliation{Departament d'Enginyeria Inform\'atica i Matem\'atiques, Universitat Rovira I Virgili, Tarragona, Spain} 

\author{A. Arenas}
\affiliation{Departament d'Enginyeria Inform\'atica i Matem\'atiques, Universitat Rovira I Virgili, Tarragona, Spain}

%% The \maketitle command is necessary to build the title page.
\maketitle

\section{Navigation Strategies in Interconnected Networks}

In the following subsection we will describe four representative random walk processes -- covering a wide variety of real physical processes --  and we will provide the corresponding transition rules to build the supra-Laplacian matrix, although other type of walkers, e.g., \cite{stanley83,sinatra2011maximal}, are also possible to implement in multiplex. 

%%%%%%%%%%%%%%%%%%%%%%%%%%%%%%%%%%%%%
\vspace{0.1truecm}
\subsection{Classical random walkers} 

The classical description of random walkers on a graph (i.e., monoplex networks) is already present in \cite{wilson1972introduction,tetali1991random}, although applications to networks with complex topology are more recent \cite{noh2004random,yang2005exploring}.

In monoplex networks, the random walker has probability $1/k_{i}$ to move from vertex $i$ to vertex $j$ in the neighborhood of $i$, where $k_{i}$ indicates the degree of a vertex $i$. The direct extension of such walks to the case of multiplex networks is to consider the inter-layer connections as additional edges available in vertex $i$. It follows that the probability of moving from vertex $i$ to vertex $j$ within the same layer $\alpha$ \emph{or} to switch to the counterpart of vertex $i$ in layer $\beta$ is uniformly distributed. In such a scenario, the normalizing factor to obtain the correct probability is the total strength $s_{i,\alpha}+S_{i,\alpha}$ of vertex $i$. The resulting transition rules for this classical random walker in a multiplex (RWC) are given in Table~\ref{tab:RWrules}. For sake of completeness, the Laplacian matrix corresponding to this process in monoplex networks is generally referred to as the ``normalized Laplacian''.

%%%%%%%%%%%%%%%%%%%%%%%%%%%%%%%%%%%%%
\vspace{0.1truecm}
\subsection{Diffusive random walkers} 
In monoplex networks, this type of random walk has been studied in detail in \cite{samukhin2008laplacian}. Here, at microscopic level, the random walker moves from a vertex $i$ to one of its neighbor with hopping rate which depends on $i$. In fact, if $s_{\max}=\max\limits_{i,\alpha}\{s_{i,\alpha}+S_{i,\alpha}\}$ is the maximum vertex strength in the network, the walker is allowed to wait in vertex $i$ with rate $1-s_{i}/s_{\max}$ and to jump to any other vertex with rate $s_{i}/s_{\max}$. Hence, the nature of this walk is very different from the classical one previously described, where the hopping rate does not depend on the vertex, and it can be shown that the corresponding Laplacian matrix, once unnormalized, is the same of a classical diffusive process (we refer to \cite{samukhin2008laplacian} for further detail).

We extend this walk to the case of multiplex networks by considering inter-layer connections as additional edges to estimate the maximum vertex strength. The resulting transition rules for this random walker in a multiplex (RWD) are given in Tab.\,\ref{tab:RWrules}.

\vspace{0.1truecm}
\subsection{Physical random walkers} 
Here we propose a new type of random walk dynamics in the multiplex, which reduces to the classical random walk in the case of monoplex. The transition rules are the same, except that we assume that the time scale to switch layer is negligible with respect to the time scale required to move from a vertex to another one in its neighborhood. Therefore, in the same time step the random walker is allowed to switch layer and to jump to another vertex, with layer-switching and the vertex-jumping actions being independent. This is a fundamental difference with the random walkers described so far, because they were not allowed to switch and jump in the same time unit. Moreover, another major difference lies in treating inter-layer connections as another type of edges, not competing with the intra-layer edges.

As an example of this dynamics, one might imagine the case of online social networks where each layer corresponds to a different social structure (e.g., Facebook and Twitter) and users play the role of vertices. In this case, the time required to a user to switch from one layer to the other one requires less than a few seconds. 

The resulting transition rules for this physical random walker in a multiplex (RWP) are given in Tab.\,\ref{tab:RWrules}. It is straightforward to show that this process is equivalent to the classical random walker in the case of monoplexes.

%%%%%%%%%%%%%%%%%%%%%%%%%%%%%%%%%%%%%
\vspace{0.1truecm}
\subsection{Maximal entropy random walkers} 
In classical random walks, a walker jumps from a vertex to a neighbor with uniform probability which depends only on the local structure, namely the vertex strength. However, it has been recently proposed a walk dynamics where the transition rate of jumps is influenced by the global structure of the network \cite{burda2009localization}, even only in presence of local information \cite{sinatra2011maximal}. More specifically, the walkers choose the next vertex to jump into maximizing the entropy of their path at a global level, whereas classical random walkers maximize the entropy of their path at neighborhood level. To achieve such maximal entropy paths, the transition rates are governed by the largest eigenvalue of the adjacency matrix and the components of the corresponding eigenvector \cite{burda2009localization}.

\begin{table}[!t]
\caption{\label{tab:RWrules}\textbf{Transition probability for four different random walk processes on multiplex.} We account for jumping between vertices (latin letters) and switching between layers (greek letters). When appearing in pairs, $j\neq i$ and $\beta\neq \alpha$ must be considered. See text for further detail.}
\scalebox{0.84}{
\begin{tabular}{ccccc}
\hline\hline
\textbf{Tr.} & \textbf{RWC} & \textbf{RWD} & \textbf{RWP} & \textbf{RWME}\\
\hline
$\mathcal{P}_{ii}^{\alpha\alpha}$ & 
$\frac{D^{\alpha\alpha}_{(i)}}{s_{i,\alpha}+S_{i,\alpha}}$ &
$\frac{s_{\max}+D^{\alpha\alpha}_{(i)}-s_{i,\alpha}-S_{i,\alpha}}{s_{\max}}$ & 
$0$ & 
$\frac{D^{\alpha\alpha}_{(i)}}{\lambda_{\max}}$\\
\hline
$\mathcal{P}_{ii}^{\alpha\beta}$  & 
$\frac{D^{\alpha\beta}_{(i)}}{s_{i,\alpha}+S_{i,\alpha}}$ &
$\frac{D^{\alpha\beta}_{(i)}}{s_{\max}}$&  
$0$ & 
$\frac{D^{\alpha\beta}_{(i)}}{\lambda_{\max}}\frac{\psi_{(\beta-1)N+i}}{\psi_{(\alpha-1)N+i}}$\\
\hline
$\mathcal{P}_{ij}^{\alpha\alpha}$  &
$\frac{W^{(\alpha)}_{ij}}{s_{i,\alpha}+S_{i,\alpha}}$ &
$\frac{W^{(\alpha)}_{ij}}{s_{\max}}$&
$\frac{W^{(\alpha)}_{ij}}{s_{i,\alpha}}\frac{D^{\alpha\alpha}_{(i)}}{S_{i,\alpha}}$ & 
$\frac{W^{(\alpha)}_{ij}}{\lambda_{\max}}\frac{\psi_{(\alpha-1)N+j}}{\psi_{(\alpha-1)N+i}}$\\
\hline
$\mathcal{P}_{ij}^{\alpha\beta}$  &
0&
0& 
$\frac{W^{(\beta)}_{ij}}{s_{i,\beta}}\frac{D^{\alpha\beta}_{(i)}}{S_{i,\alpha}}$ & 
0\\
\hline\hline
\end{tabular}
}
\end{table}

In the case of multiplex, we use the supra-adjacency matrix 
\begin{eqnarray}
\mathcal{A}=\left(
\begin{array}{c|c|c|c}
  D^{11}\mathbf{I}+\mathbf{W}^{(1)} & D^{12}\mathbf{I} & \dots & D^{1L}\mathbf{I} \\ \hline
  D^{21}\mathbf{I} & D^{22}\mathbf{I}+\mathbf{W}^{(2)} & \dots & D^{2L}\mathbf{I} \\ \hline
  \vdots & &\ddots &  \vdots\\\hline
  D^{L1}\mathbf{I} &   D^{L2}\mathbf{I} & \dots & D^{LL}\mathbf{I}+\mathbf{W}^{(L)} \\ 
\end{array}
\right)\nonumber
\end{eqnarray}
to achieve the same result (see Materials and Methods in the main text for further detail). We indicate with $\lambda_{\max}$ the largest eigenvalue of this matrix and with $\mathbf{\psi}$ the corresponding eigenvector. Therefore, according to the prescription given in \cite{burda2009localization}, the resulting transition rules for this maximal entropy random walker in a multiplex (RWME) are given in Tab.\,\ref{tab:RWrules}.

 \begin{figure*}[!t]
 \centering
 \subfigure[ BA+ER, $D^{12}=D^{21}=1$]
   {\includegraphics[width=9cm]{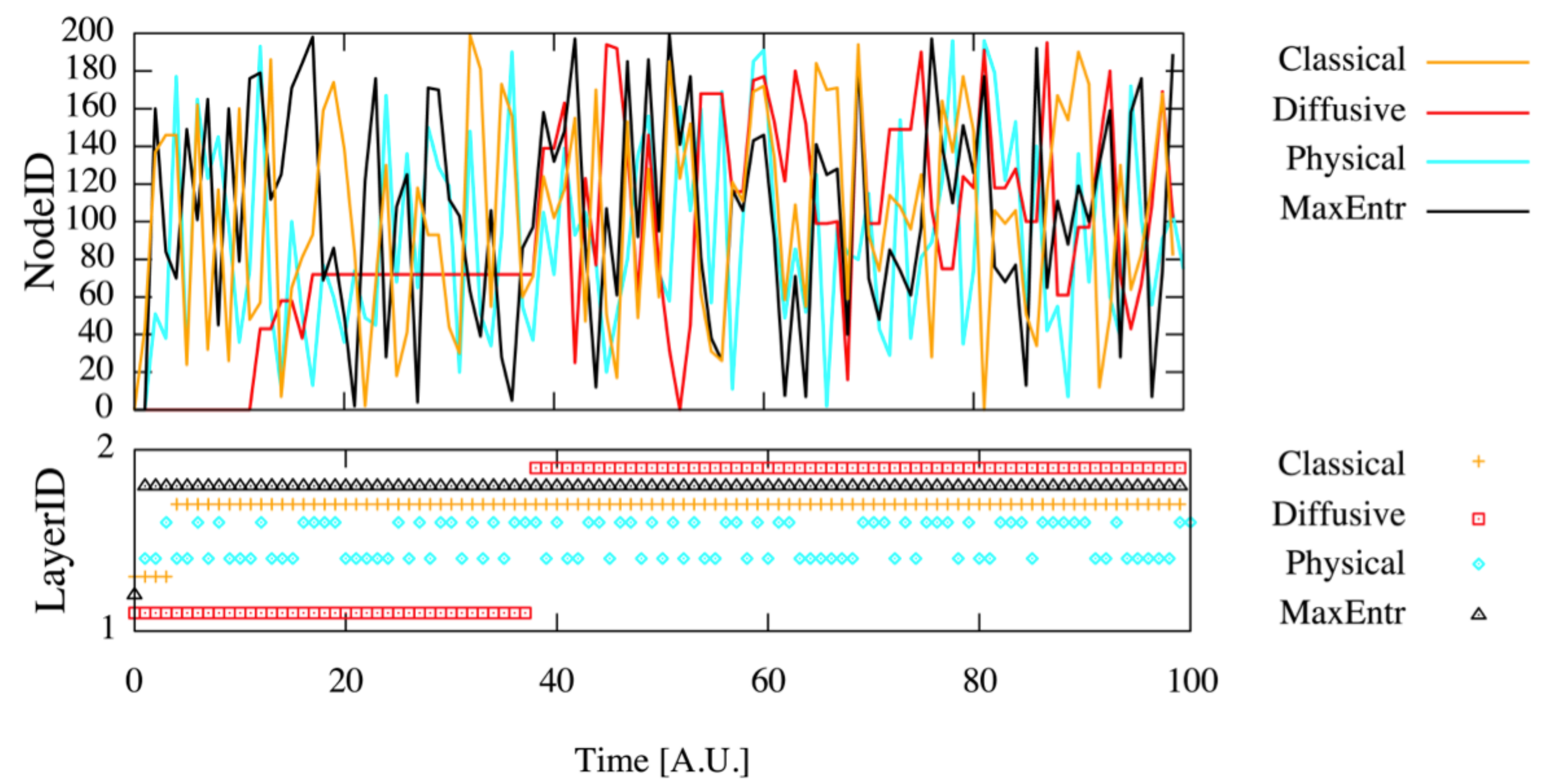}}
 \hspace{-3mm}
 \subfigure[ BA+ER, $D^{12}=D^{21}=100$]
   {\includegraphics[width=9cm]{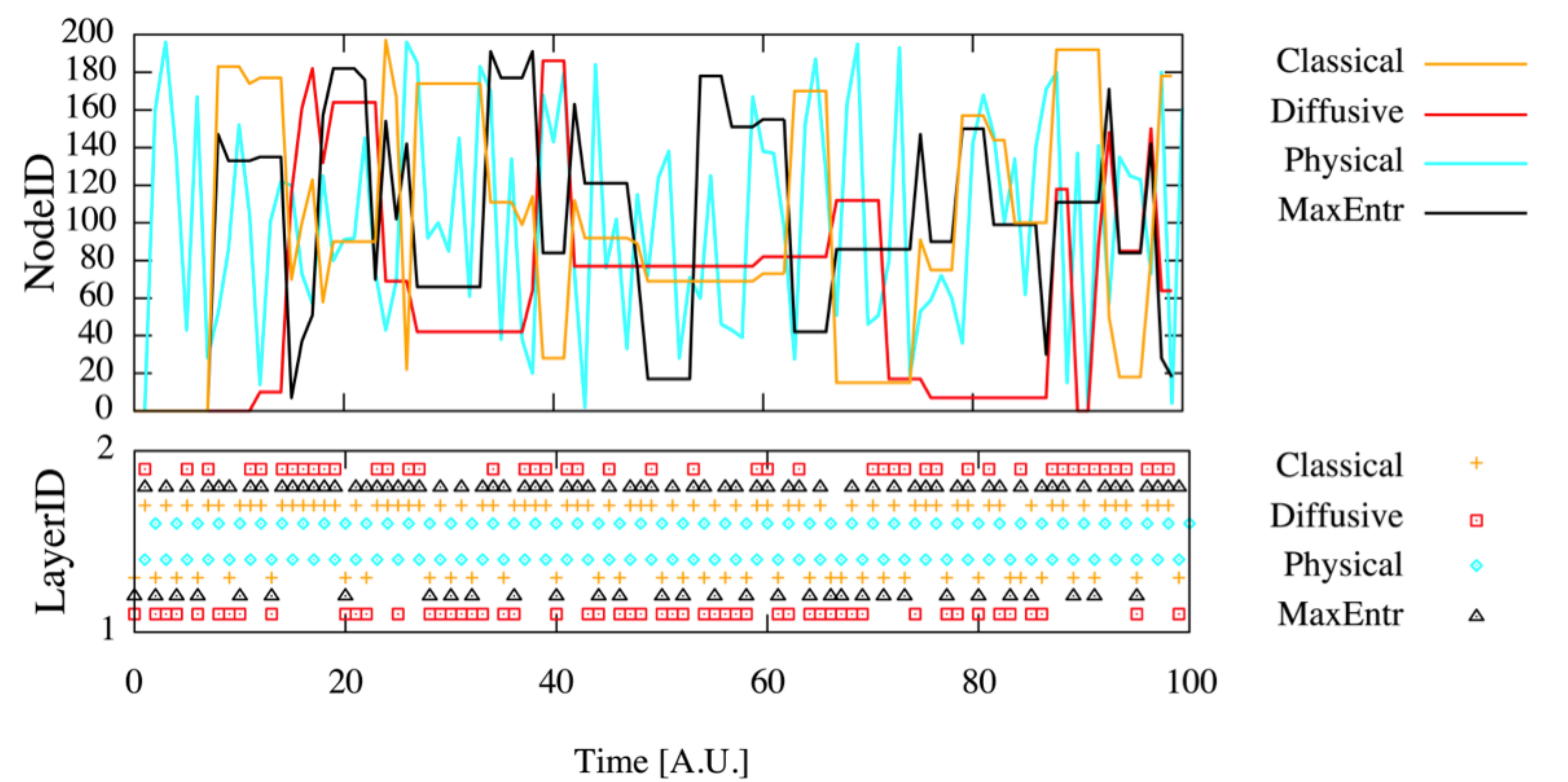}}
 \caption{\label{fig:rw-traj-comp}\textbf{Random walks realizations on different multiplex structures.} Vertices (top panels) and layers (bottom panels) visited by one random walker in 100 time steps. The four types of walk considered in this study are shown. The multiplex is built with one Barab{\'a}si-Albert (layer one) and one Erd{\H{o}}s-R{\'e}nyi (layer two) network with 200 vertices, while inter-layer weights are specified above.}
 \end{figure*}
 
A representative example of each walk is shown in \Fig{fig:rw-traj-comp}, where vertices and layers visited by one random walker up to 100 time steps are reported. We show two different cases, corresponding to different choices of inter-layer weights, to make evident the difference in the dynamics.

\begin{figure*}[!t]
\centering
\small\addtolength{\tabcolsep}{-2mm}
\begin{tabular}{cccc}
\includegraphics[width=4.45cm]{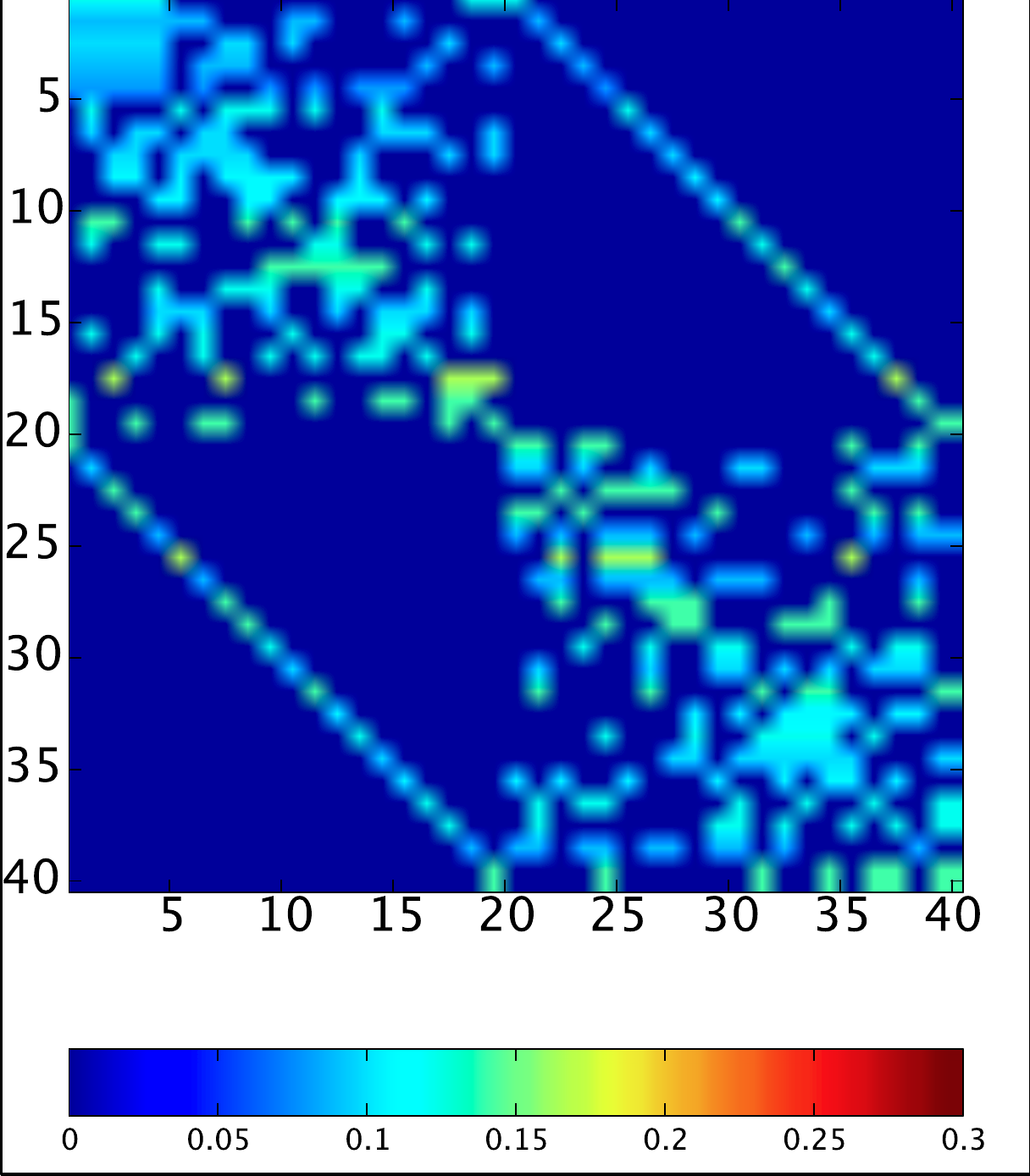} & 
\includegraphics[width=4.45cm]{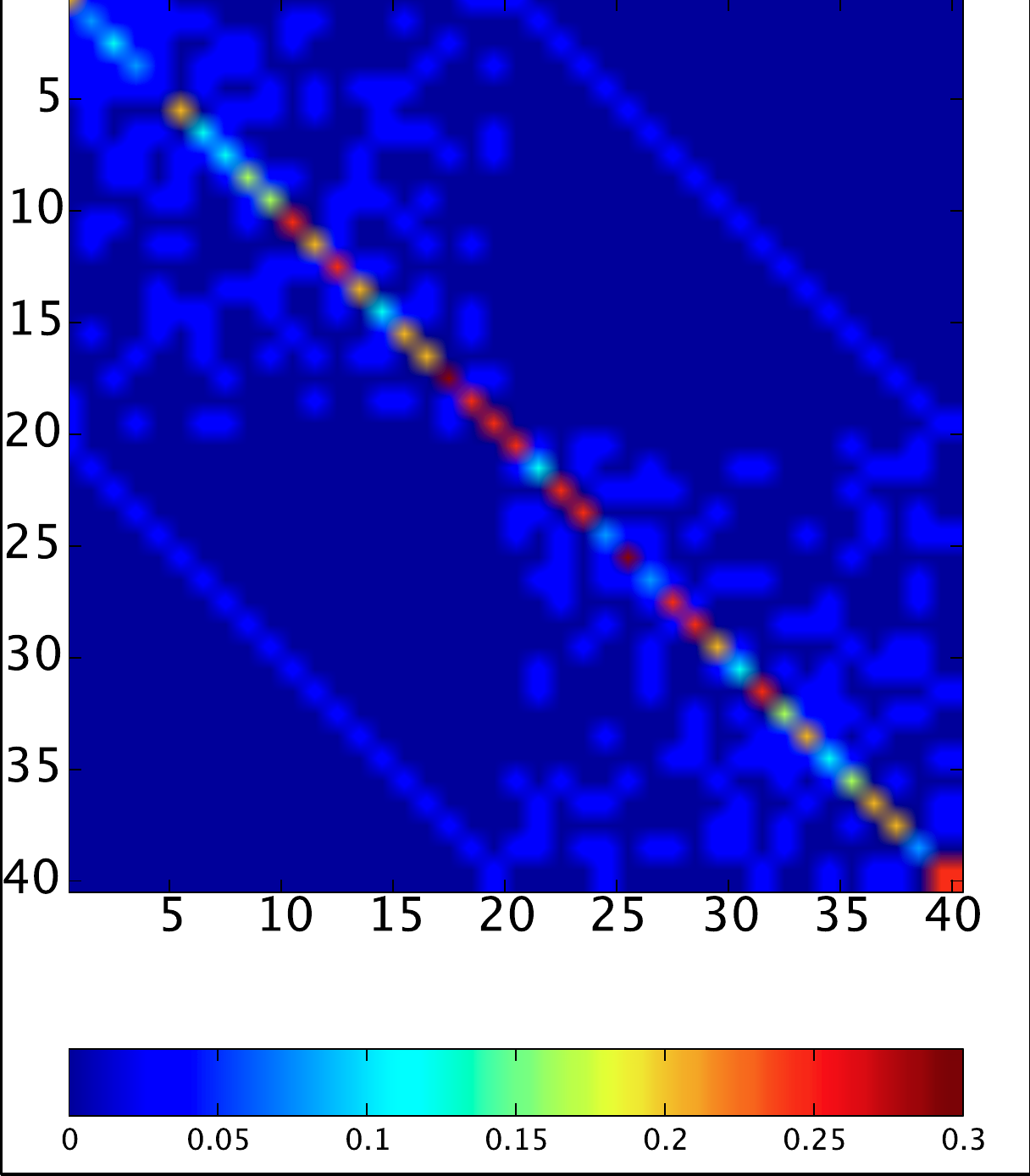} &
\includegraphics[width=4.45cm]{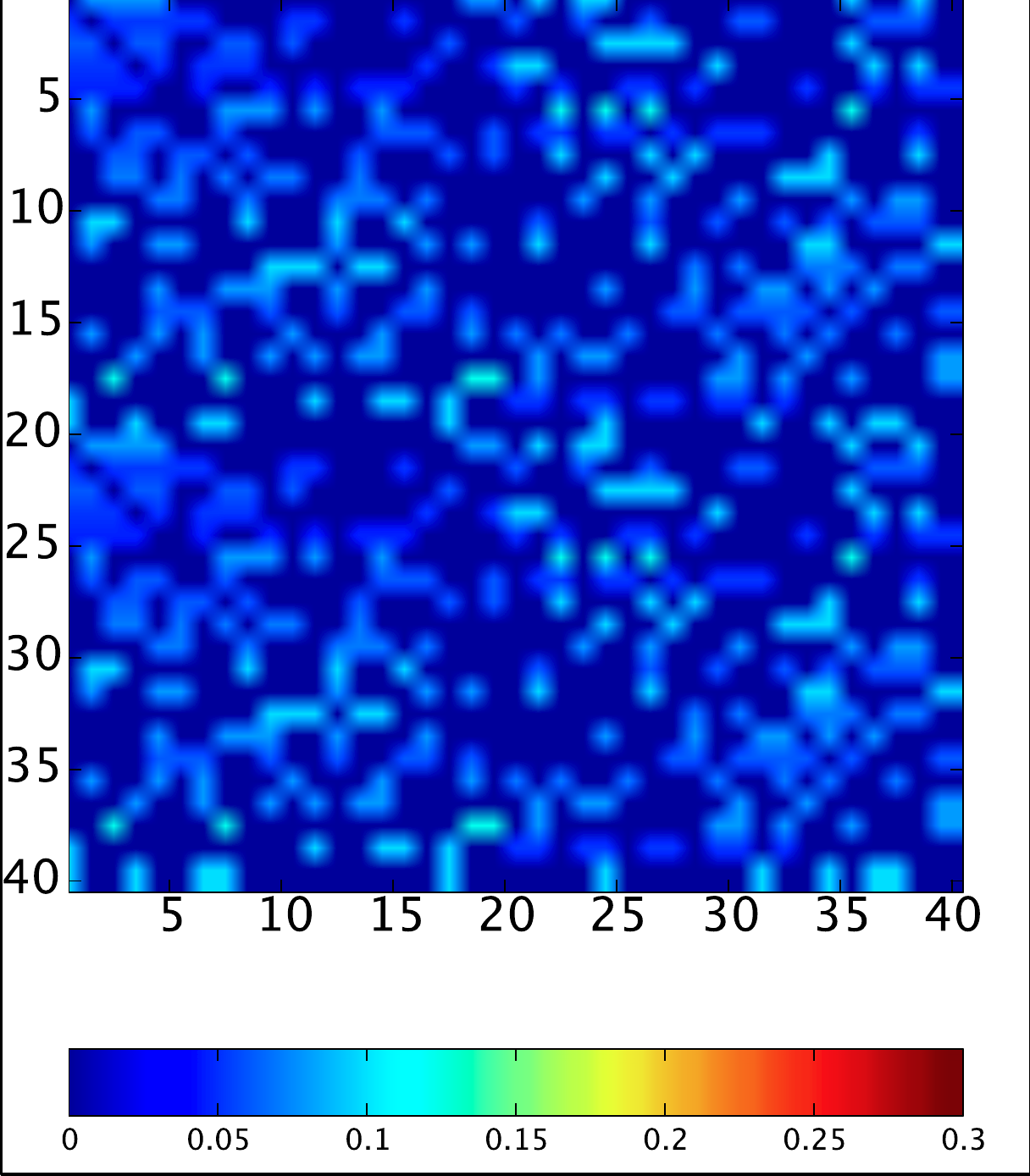} &
\includegraphics[width=4.45cm]{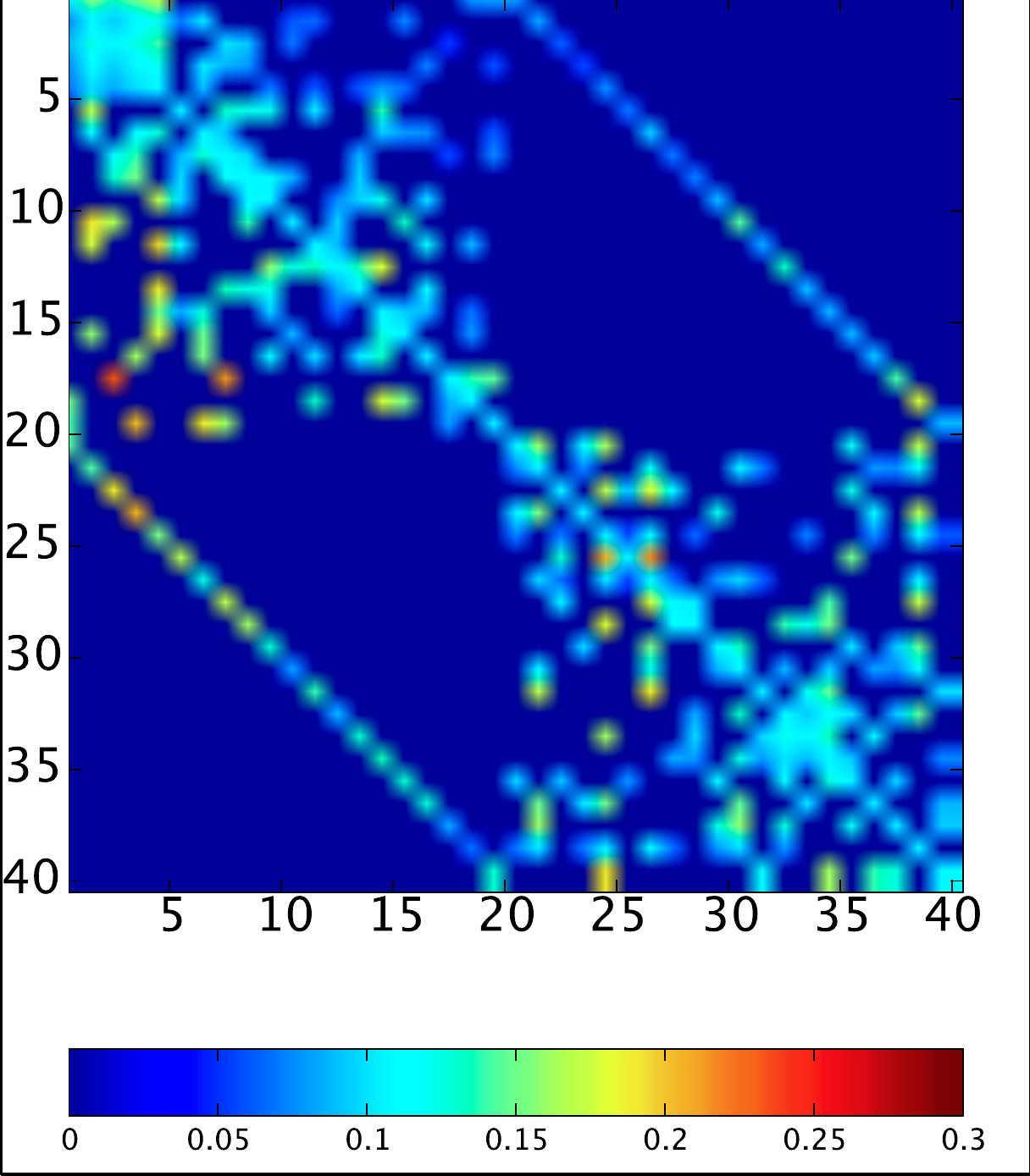}\\
\includegraphics[width=4.45cm]{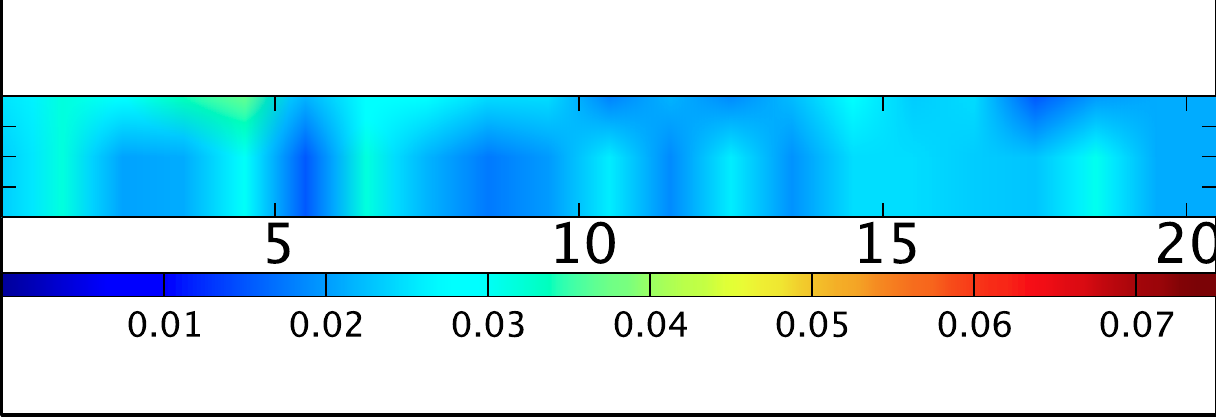} & 
\includegraphics[width=4.45cm]{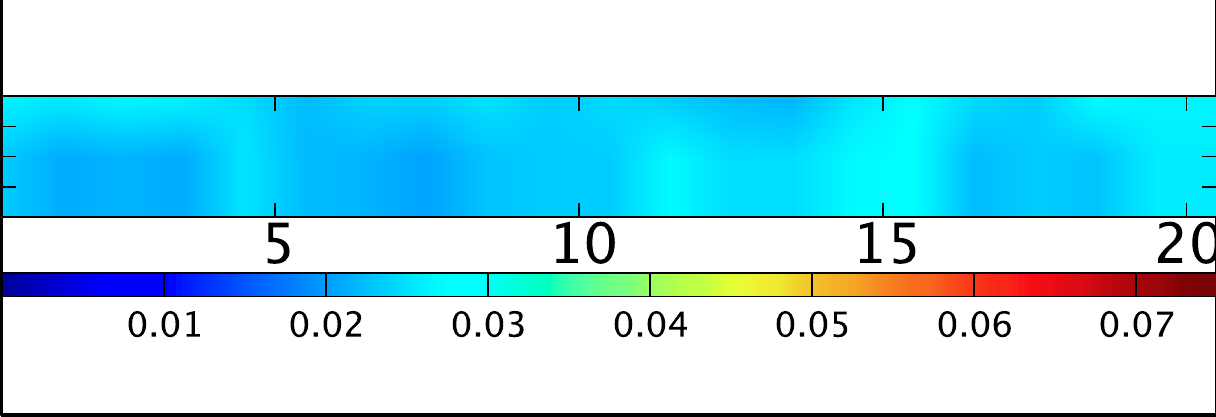} &
\includegraphics[width=4.45cm]{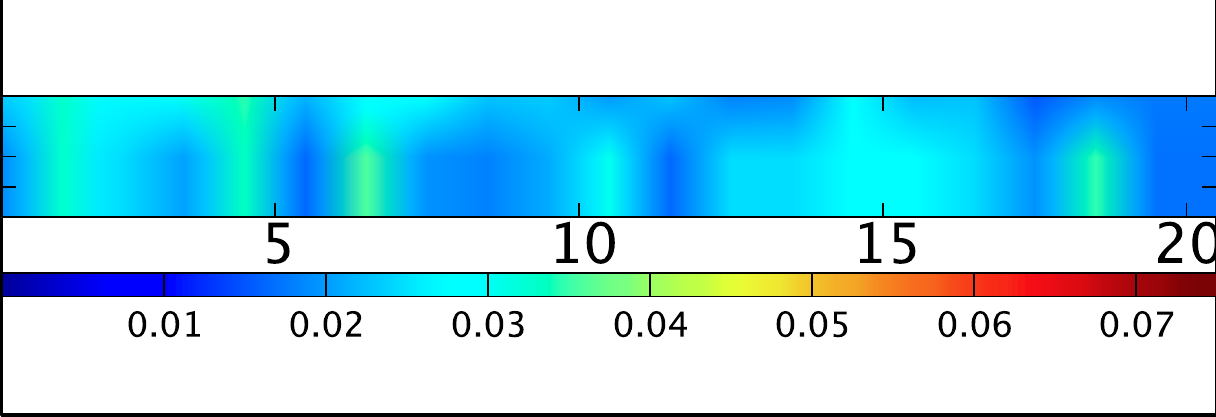} &
\includegraphics[width=4.45cm]{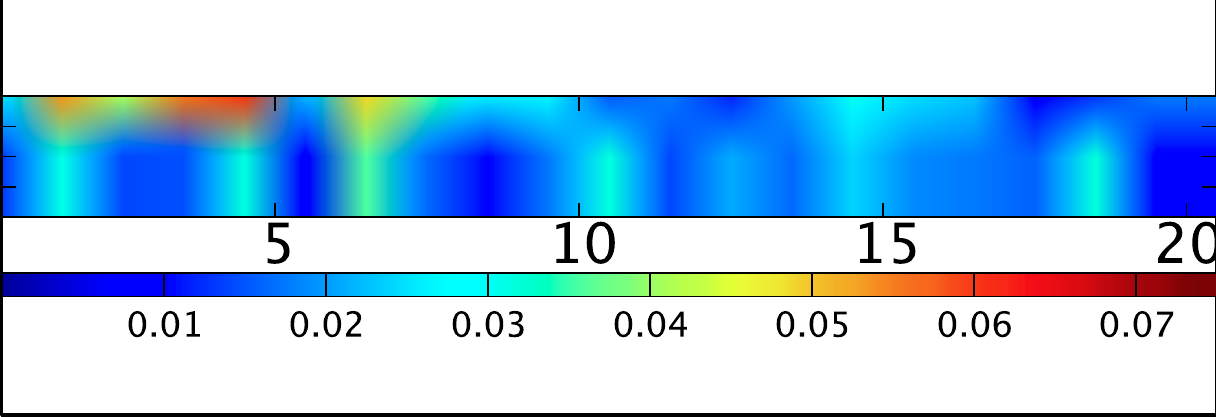}\\
\includegraphics[width=4.45cm]{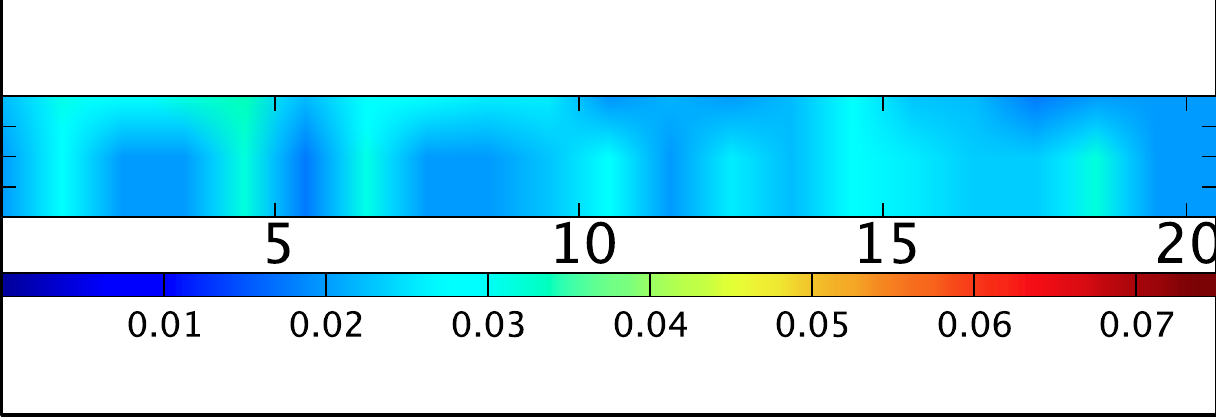} & 
\includegraphics[width=4.45cm]{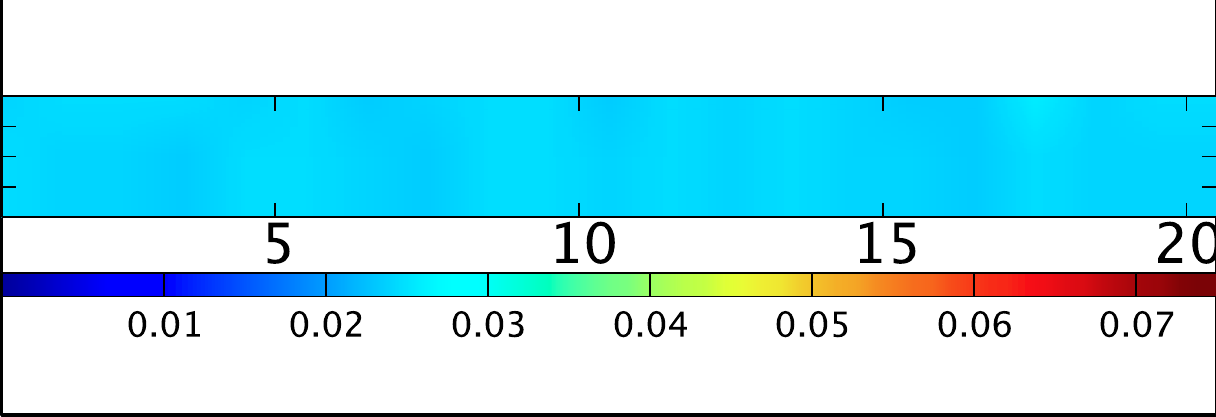} &
\includegraphics[width=4.45cm]{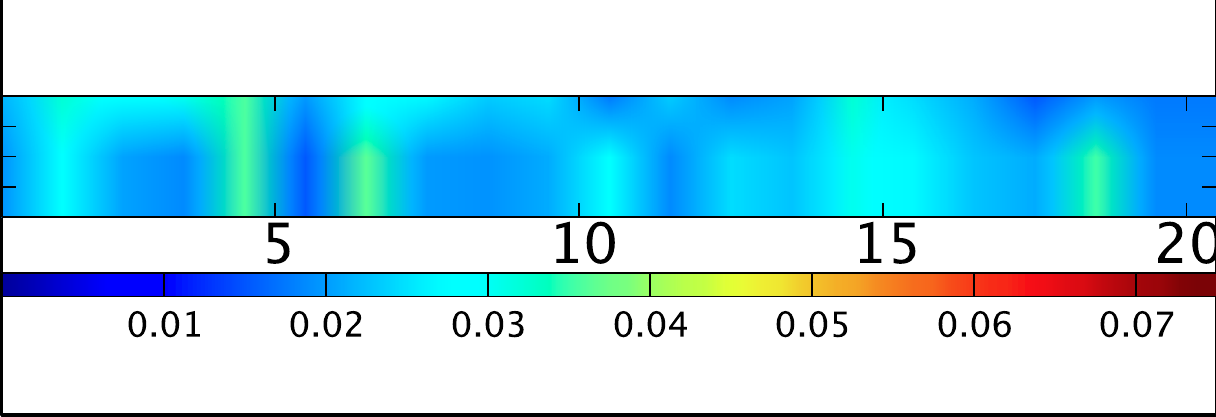} &
\includegraphics[width=4.45cm]{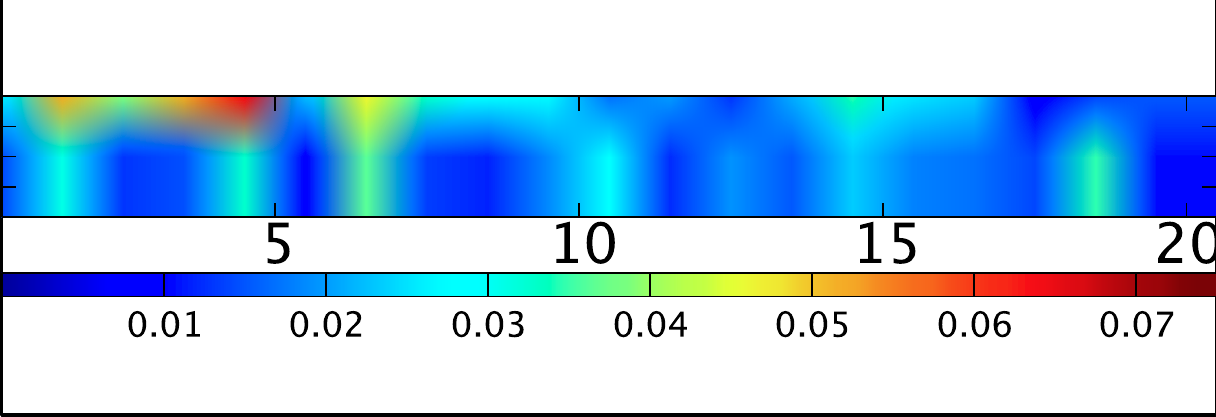}\\
Classical & Diffusive & Physical & Maximal entropy
\end{tabular}
\caption{\label{fig:rw-panels}\textbf{Probabilities governing four random walk strategies on multiplex.} \emph{Top panels:} transition probabilities for walks considered in this study. Note that we have rescaled by a factor 2 the transition matrix of diffusive walk for better visualization and to allow comparisons. \emph{Middle panels:} occupation probability, for each vertex in each layer, considering one random walk starting only from the first vertex. \emph{Bottom panels:} as in middle panels, but considering one random walk starting with uniform probability from any other vertex. Multiplex of 20 vertices embedded in two different realizations of a Watts-Strogatz small-world network (rewiring probability is $0.2$), where $D^{11}=D^{21}=D^{12}=D^{21}=1$. Different exploration strategies are responsible for the different probability that a vertex is visited and occupied by a random walker.}
\end{figure*}

In the top panels of Fig.\,\ref{fig:rw-panels} we show the transition probabilities in the case of a multiplex of 20 vertices embedded in two different realizations of a Watts-Strogatz small-world network \cite{watts1998collective}. The probability to find a random walker in a certain vertex on a certain layer is also shown in the same figure, considering one walk starting from the first vertex only (middle panels) and from any other vertex with uniform probability (bottom panels). As expected, different exploration strategies result in different occupation probability, where some vertices in a certain layer might be explored more (or less) frequently, as in the case of RWC, RWP and RWME, or uniformly as in the case of RWD. 

\Fig{fig:rw-traj-comp} and \Fig{fig:rw-panels} clearly highlight the different dynamics and how navigation strategy influences the exploration of the multiplex.

%%%%%%%%%%%%%%%%%%%%%%%%%%%%%%%

\section{Occupation Probability of Random Walkers}

We define the \emph{occupation probability} $\Pi_{i,\alpha}=\lim\limits_{t\lto\infty}p_{i,\alpha}(t)$ to find a walker in vertex $i$ of layer $\alpha$ in the limit $t\lto \infty$, and we indicate with $\mathbf{\Pi}$ the corresponding supra-vector. In general, $\mathbf{\Pi}$ is the left eigenvector of the supra-transition matrix corresponding to the unit eigenvalue. In some cases, the occupation probability can be estimated from the detailed balance equation
\begin{equation}
\Pi_{i,\alpha}\mathcal{P}^{\alpha\beta}_{ij}=\Pi_{j,\beta}\mathcal{P}^{\beta\alpha}_{ji}, 
\end{equation}
obtaining 
\begin{equation}
\Pi_{i,\alpha} = \frac{s_{i,\alpha}+S_{i,\alpha}}{\sum_{\beta}\sum_{j}s_{j,\beta}+S_{j,\beta}}
\end{equation}
for RWC, generalizing the well-known result obtained for walks in a monoplex network,
\begin{equation}
\Pi_{i,\alpha} = \frac{1}{NL}
\end{equation}
for RWD, as expected for a purely diffusive walk, and 
\begin{equation}
\Pi_{i,\alpha} = \psi^{2}_{(\alpha-1)N+i},
\end{equation}
for RWME, generalizing the results obtained in \cite{burda2009localization} for monoplex networks.

Indeed, following the approach proposed in \cite{noh2004random} for random walks on monoplexes, it is possible to show that the time required to a random walker starting from vertex $i$ to arrive back to the same vertex, i.e., the mean return time, is given by 
\begin{equation}
\langle T_{ii}\rangle=\frac{1}{\sum\limits_{\alpha=1}^{L}\Pi_{i,\alpha}}. 
\end{equation}
It is straightforward to verify that distributions expected in the case of monoplex are recovered for $L=1$. It is worth noting that for classical random walks the occupation probability of vertex $i$ is proportional to its \emph{supra-strength}, i.e., intra- plus inter-layer strengths, whereas for diffusive walks such a probability is the same for any vertex, regardless of multiplex topology.

%%%%%%%%%%%%%%%%%%%%%%%%%%%%%%%%%%%

\section{Dynamical vs Topological Descriptors}

\begin{figure}[!t]
 \centering
   \includegraphics[width=8.5cm]{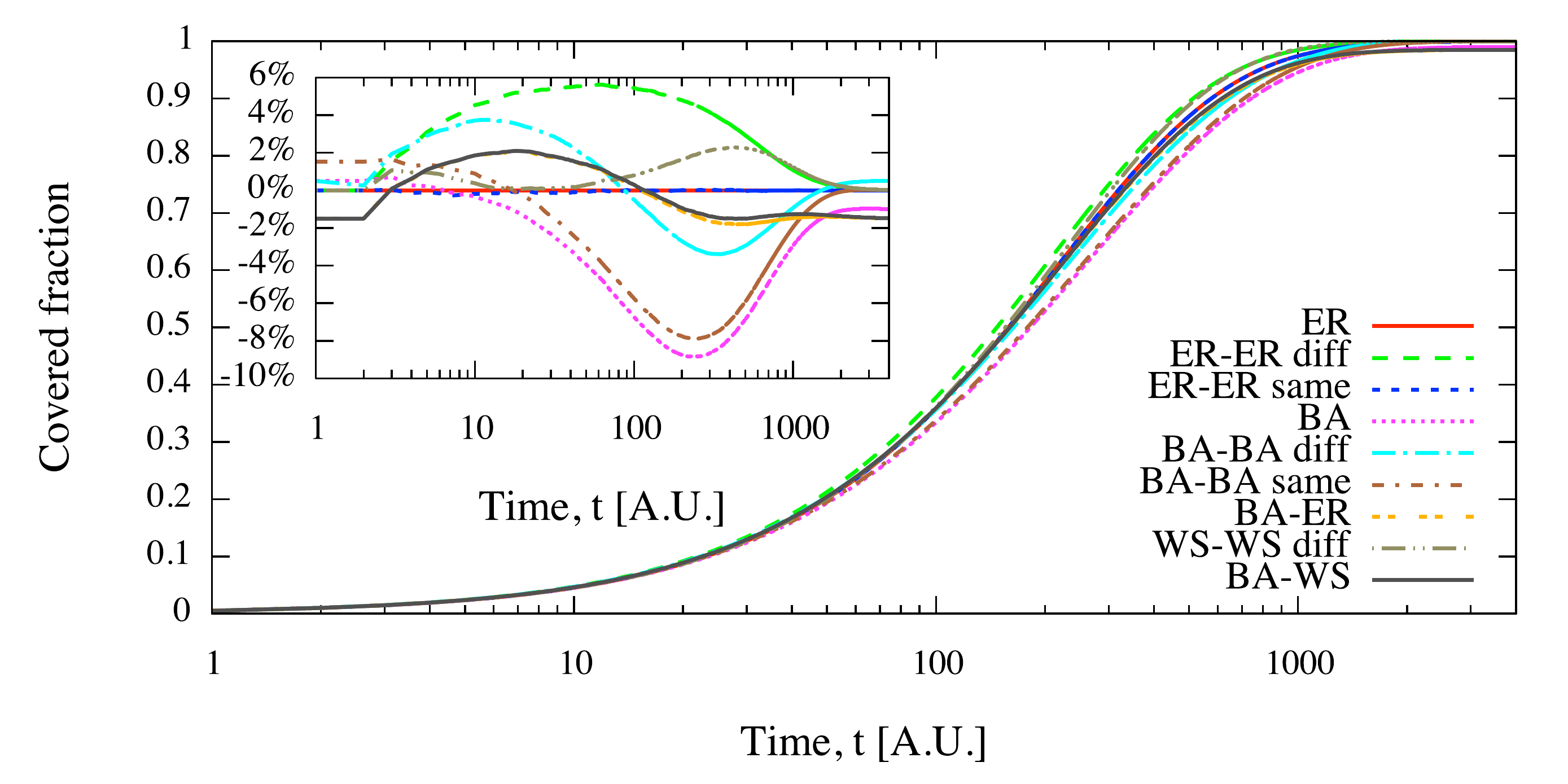}
 \caption{\label{fig:cover-fract-example}\textbf{Dependence of the coverage on multiplex topology.} Number of visited vertices \emph{versus} time for monoplex and multiplex topologies (see the text for further details about the simulations). The inset shows the relative difference of each curve with respect to the coverage obtained for an ER monoplex, evidencing that vertices in different topologies are visited with different time scales.}
\end{figure}

We show in \Fig{fig:cover-fract-example} the coverage \emph{versus} time in the case of RWP only, for some representative multiplexes where $D^{12}_{(i)}=D^{21}_{(i)}=D^{11}_{(i)}=D^{22}_{(i)}=1$, $\forall i=1,2,...,N$. Results for different combination of topologies (double acronym in the legend) are shown, together with results for walks in a single layer (single acronym in the legend). ``Diff'' indicates same topology but different random realizations, while ``same'' indicates same topology and same random realization on both layers. Inset shows the relative difference of coverages with respect to the case of an ER monoplex. 

The multiplex topology has an evident impact on the walk process, delaying or accelerating the exploration of the network with respect to a random search in a monoplex random network.

\begin{figure}[!t]
 \centering
   \includegraphics[width=8.5cm]{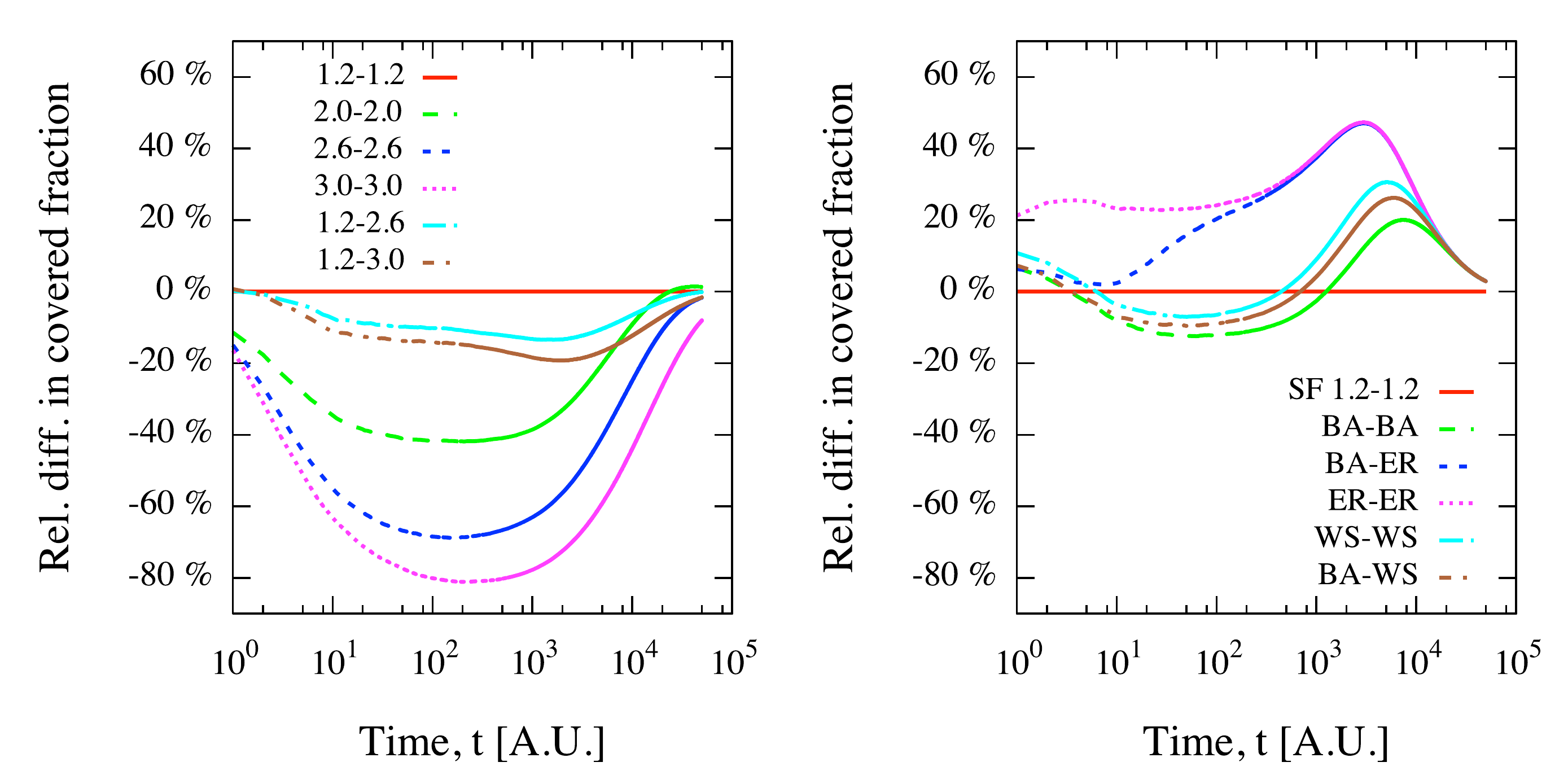}
   \includegraphics[width=8.5cm]{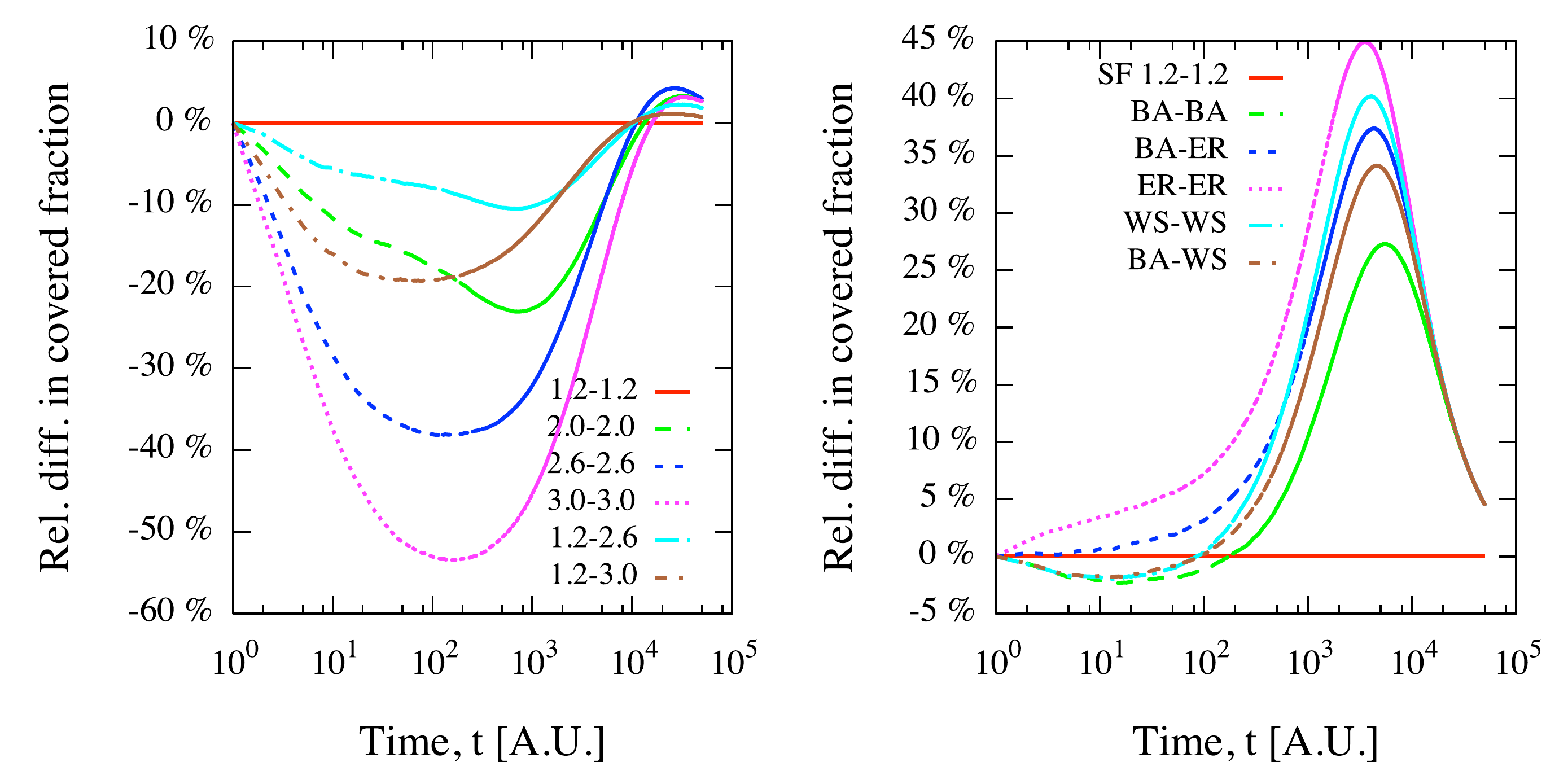}
 \caption{\label{fig:cover-fract-example2}\textbf{Dependence of the coverage on multiplex topology.} Same as the inset of \Fig{fig:cover-fract-example2}, where the relative difference of each curve is calculated with respect to the coverage obtained for a multiplex of two different scale-free networks with degree distribution $\propto k^{-1.2}$. Top panels refer to RWC, whereas bottom panels refer to RWP. Left panels (top and bottom) refer to multiplexes of different scale-free networks with other degree distributions, show indices are specified in the legend. Right panels (top and bottom) refer to multiplexes of other topologies.}
\end{figure}

This is a genuine effect of the multi-layer structure and it is not related to the finite size of the considered networks. In fact, as shown in \Fig{fig:cover-fract-example2}, where multiplexes of 2000 nodes and many different topologies are considered.

\begin{figure}[!t]
	\centering
	  \includegraphics[width=8.7cm]{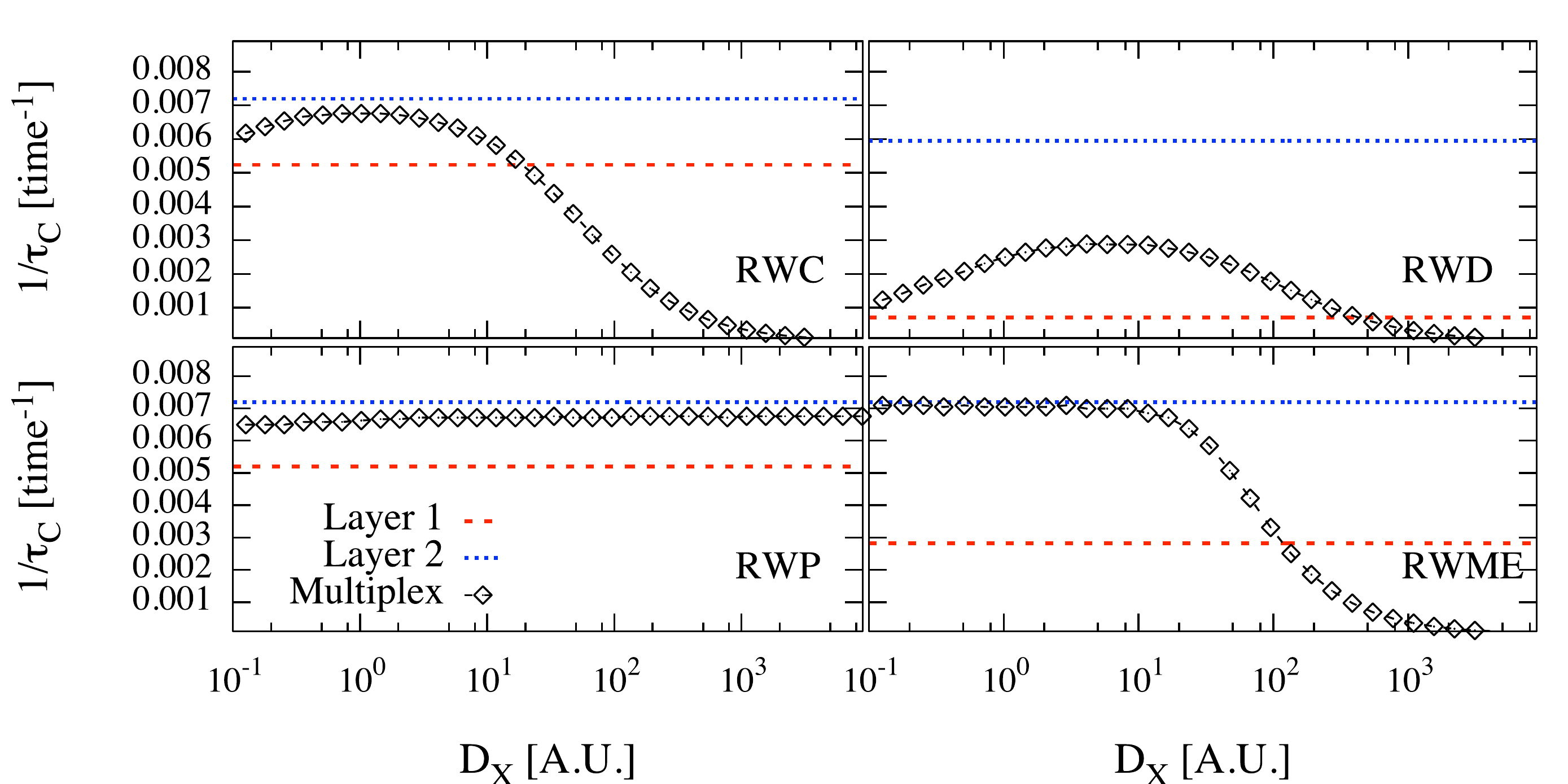}
	\caption{\label{fig:aber-invcov-comp}\textbf{Critical dependence of the coverage on navigation strategy and inter-layer connection strength.} Different random walks are used to calculate the inverse of the time $\tau_{C}$ required to cover the 50\% of a BA+ER multiplex with 200 vertices, as a function of $D_{X}=D^{12}=D^{21}$. The values for walks in each layer are shown for comparison and make clear how different exploration strategies have a strong effect on the coverage time scale.}
\end{figure}

\begin{figure}[!t]
 \centering
   \includegraphics[width=8.5cm]{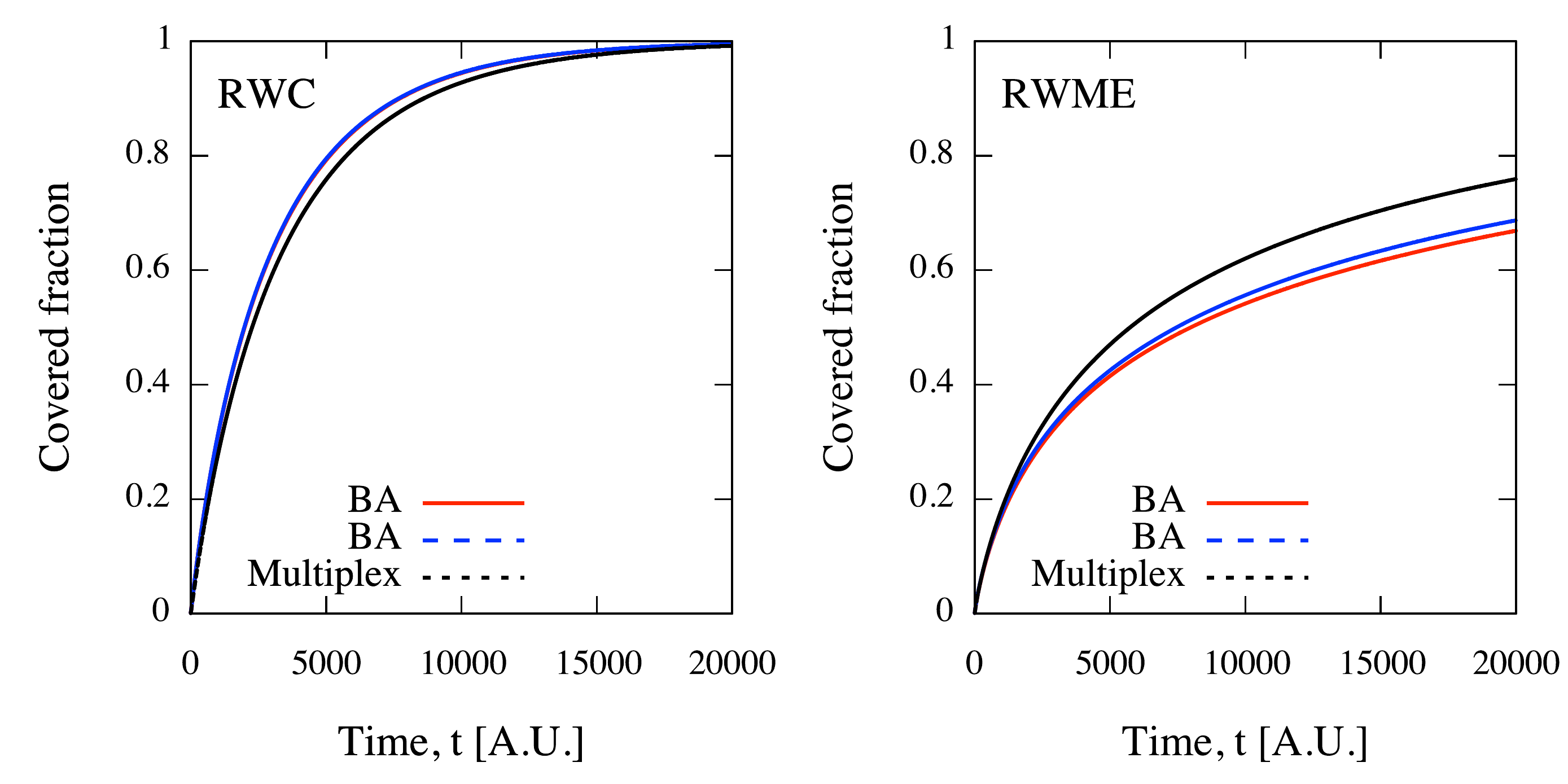}
   \includegraphics[width=8.5cm]{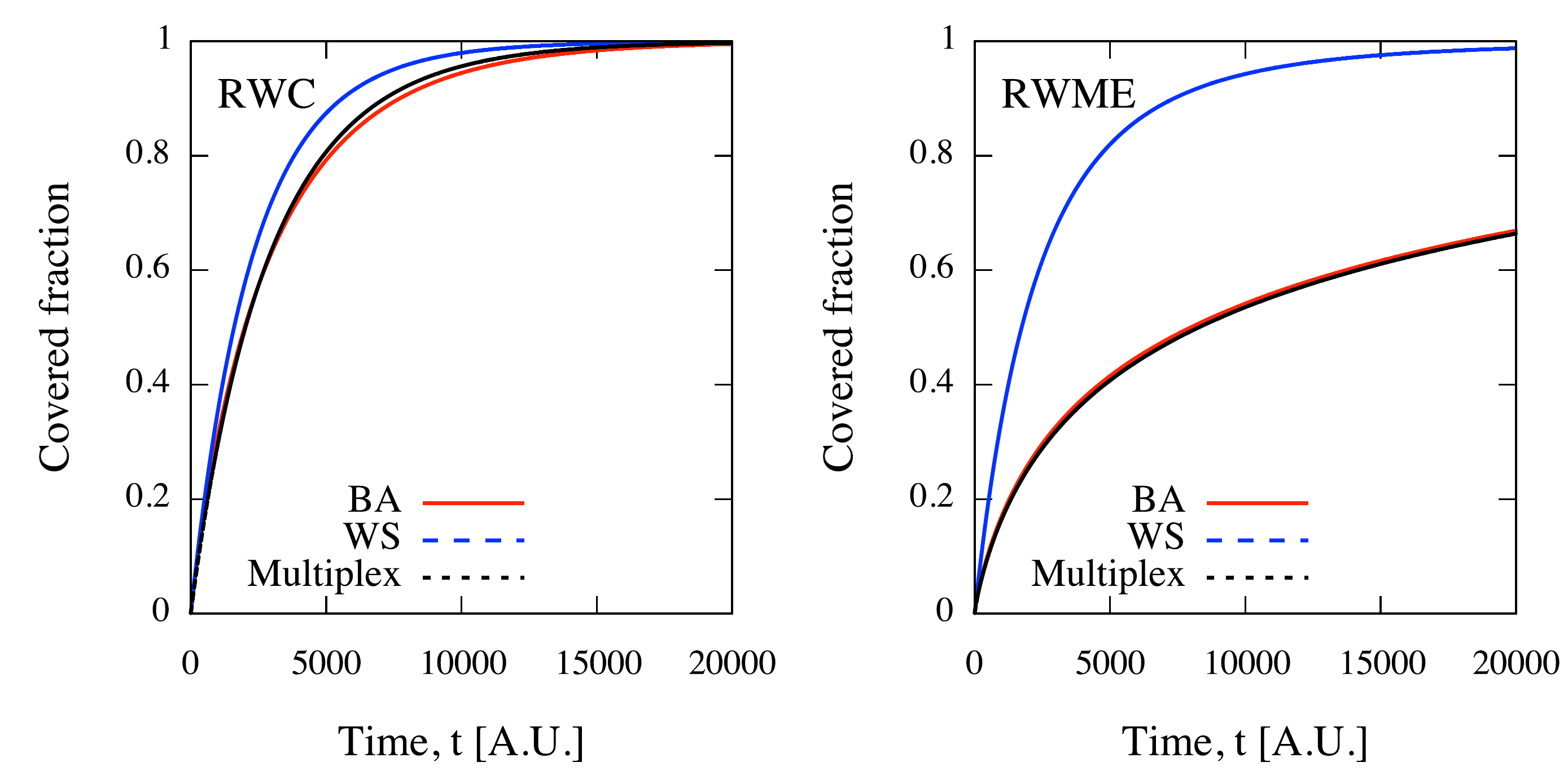}
 \caption{\label{fig:cover-diff-type}\textbf{Different types of diffusion characterize different topological structures and navigation strategies.} Coverage \emph{versus} time for two different multiplex topologies (BA+BA on the top panels and BA+WS on the bottom panels) and two different walk rules (RWC on the left panels and RWME on the right panels). While the diffusion on single layers separately and on the multiplex is similar for RWC on BA+BA, this is not the case for RWME on BA+BA where enhanced diffusion is shown in the multiplex. In the other cases, the diffusion is \emph{infra}-diffusive.}
\end{figure}

In \Fig{fig:aber-invcov-comp}, for each random walk considered, we show the inverse of the time $\tau_{C}$ required to cover the 50\% of a BA+ER multiplex with 200 vertices as a function of the inter-layer weight $D_{X}=D^{12}=D^{21}$. It is worth mentioning that the final result depends only quantitatively, but not qualitatively, on the choice of the covered fraction. This representative example shows the impact of transition rules on the exploration of the multiplex, putting in evidence that the best strategy to adopt to cover the network depends on the topology and on the weight of inter-layer connections. Moreover, in this specific experiment, the walk in the multiplex is \emph{infra-diffusive} (\emph{sub-diffusive}) depending on the value of $D_{X}$, i.e., the time to cover the multiplex lies between (is smaller than) the times required to cover each layer separately. It is worth noting that in other cases, like  RWME on BA+BA multiplexes, walks show \emph{enhanced diffusion}, i.e., the time to cover the multiplex is smaller than the time to cover each layer separately. This is shown, for instance, in \Fig{fig:cover-diff-type}.

\begin{figure}[!t]
 %\vspace{-5mm}
 \centering
   \includegraphics[width=8.5cm]{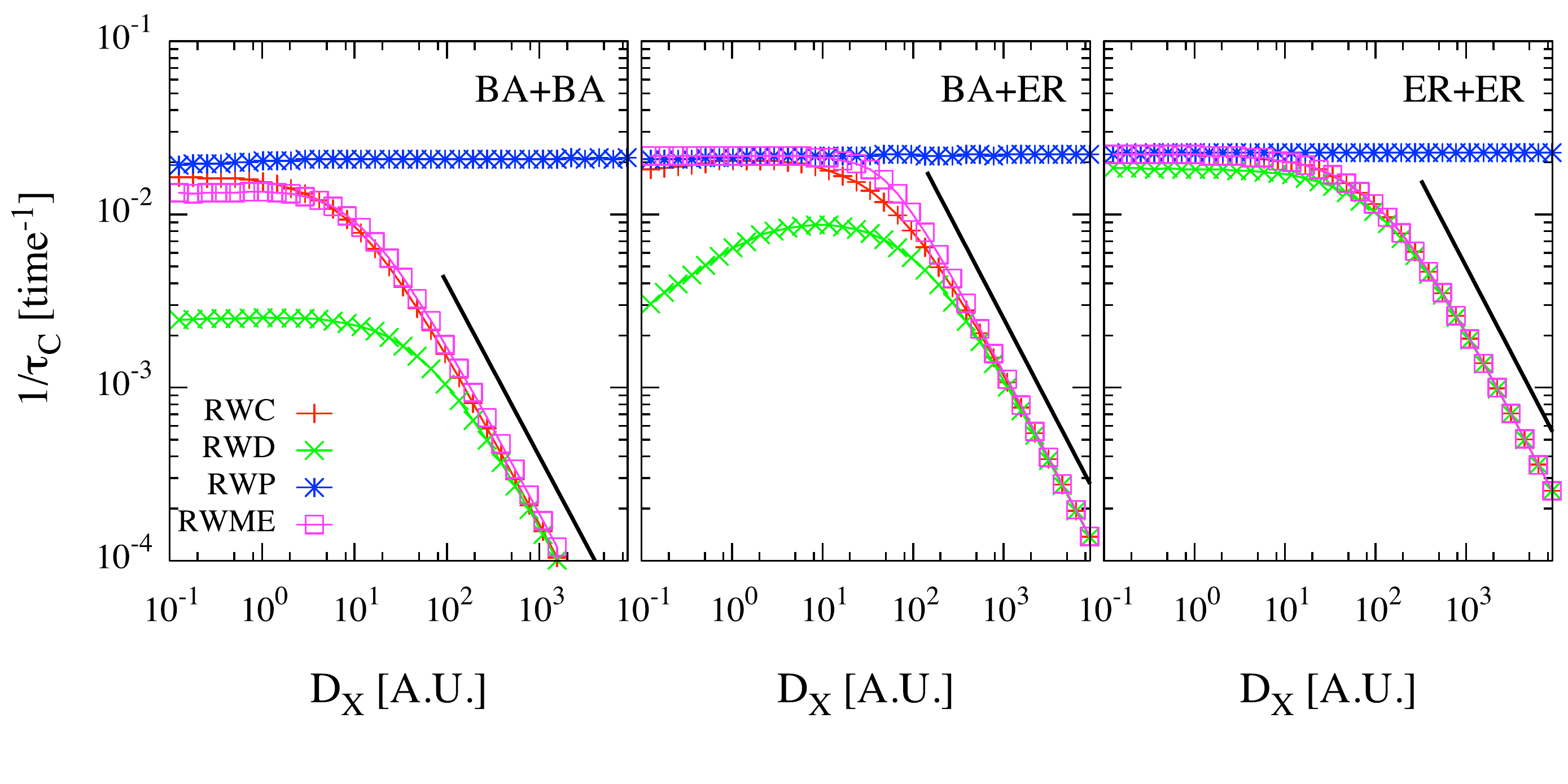}
   \includegraphics[width=8.5cm]{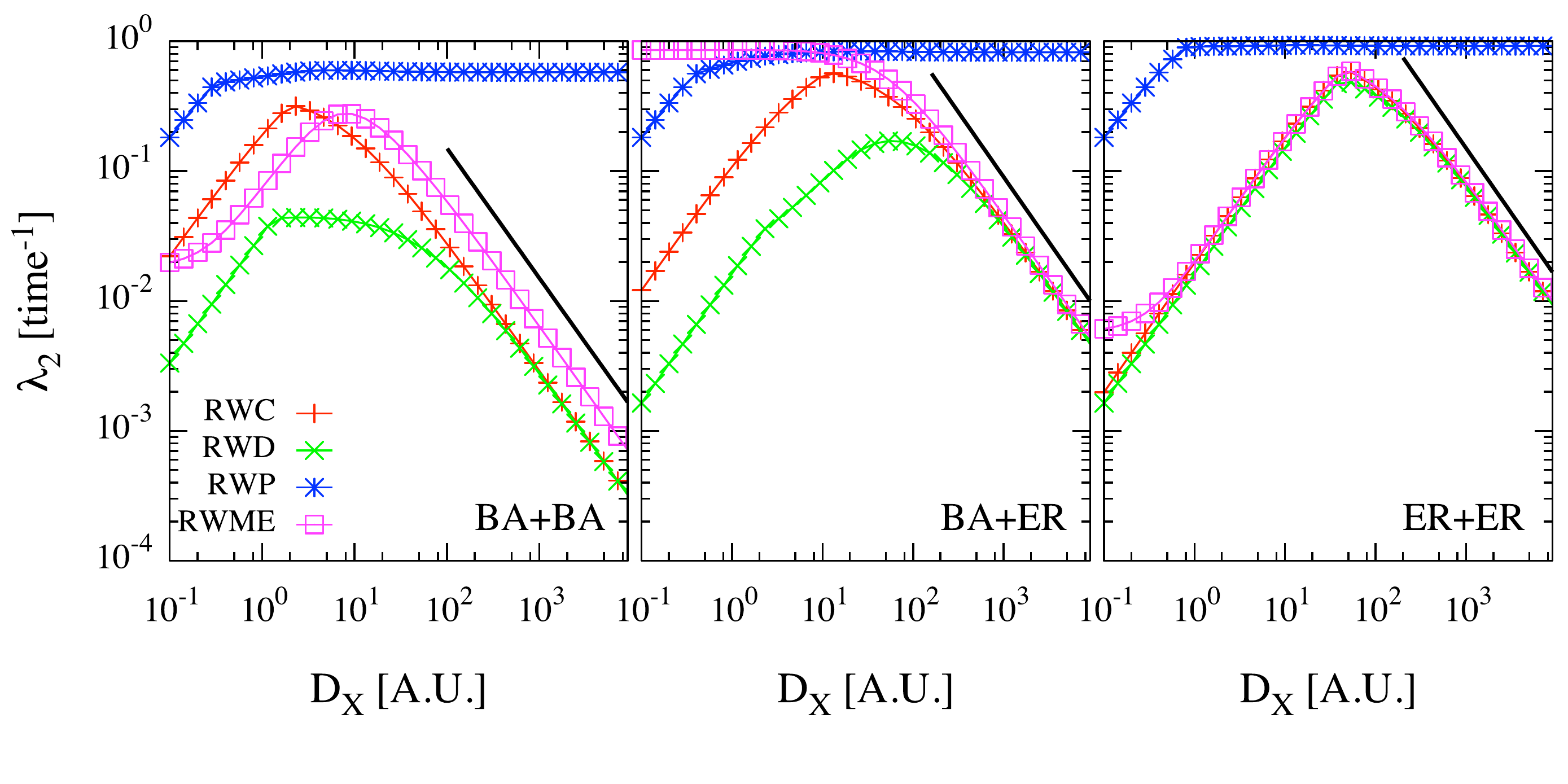}
 \caption{\label{fig:invcov-vs-lambda2}\textbf{Relation between dynamical and topological descriptors of a multiplex.} Inverse of the time required to cover 50\% of the network (top panels) and second smallest eigenvalue of the supra-Laplacian (bottom panels) as a function of $D_{X}$ for three different multiplex topologies and different random walk. The solid straight line indicates $D_{X}^{-1}$. These results show an intimate relationships between the structure of the multiplex and the dynamics of the stochastic process taking place on it.}
 \end{figure}

Intriguingly, we observe a similar behavior for $\lambda_{2}$, i.e., the second smallest eigenvalue of the supra-Laplacian. We show in \Fig{fig:invcov-vs-lambda2} the values of $1/\tau_{C}$ (top panels) and $\lambda_{2}$ (bottom panels) versus $D_{X}$ for the four random walks and three different multiplex topologies with 200 vertices, namely BA+BA (left panels), BA+ER (middle panels) and ER+ER (right panels). Except for the smallest values of $D_{X}$, the behavior is the same, especially in the limit of $D_{X}\lto\infty$. 

See the main text and the corresponding Materials and Methods for a qualitative explanation of this result. From 
\begin{eqnarray}
\label{eq:rhovslambda2}
\rho(t)\approx1-\frac{1}{N^{2}}\sum_{i,j=1}^{N}\Delta_{ij}e^{-\mathcal{C}_{i,j}(1)t-\mathcal{C}_{i,j}(2)\lambda_{2}^{-1}},
\end{eqnarray}
the importance of $\lambda_{2}$ in the evolution of the coverage is evident. Let $\tau^{\star}$ be the time required to cover a certain fraction $\rho^{\star}=\rho(\tau^{\star})$. For large values of $\tau^{\star}$, the weighted sum of exponentials in \Eq{eq:rhovslambda2} is dominated by terms with largest temporal scale of exponential decay, i.e., by terms where the constants $\mathcal{C}_{i,j}(1)$ are the minimum ones. We indicate with $\mathcal{C}_{r,s}(1)$ the smallest among all such constants. In the worst case, all terms equally contribute to $\rho(\tau^{\star})$ and, therefore, the following inequality is satisfied:
\begin{eqnarray}
\rho^{\star}\leq 1-e^{-\mathcal{C}_{r,s}(1)\tau^{\star}-\mathcal{C}_{r,s}(2)\lambda_{2}^{-1}}.
\end{eqnarray}
A rough estimation $\tau$ of $\tau^{\star}$ can be obtained by always considering the case with equality in the above formula, leading to
\begin{eqnarray}
\label{eq:lbtau}
\tau\approx -\frac{\frac{\mathcal{C}_{r,s}(2)}{\lambda_{2}}+\log\(1-\rho^{\star}\)}{\mathcal{C}_{r,s}(1)}.
\end{eqnarray}
By using the Perron-Frobenius it is possible to show that $\mathcal{C}_{r,s}(1)\geq0$. To have a positive value of $\tau$, the numerator in \Eq{eq:lbtau} should be negative, i.e., we are able to provide an estimation only for temporal scales such that the corresponding coverage satisfies the additional constraint $\rho^{\star}>1-\exp\[-\mathcal{C}_{r,s}(2)\lambda_{2}^{-1}\]$.

From \Eq{eq:lbtau} it is evident the strong influence of $\lambda_{2}$ on the inverse coverage time. The constants $\mathcal{C}_{r,s}(1)$, playing a crucial role in the time evolution of the coverage, explicitly depend on eigenvector centralities and are smaller for more peripheral vertices which are less reachable because of the topological structure and the nature of the walk.

It is also worth investigating the behavior of  \Eq{eq:rhovslambda2} in the limit of small or large values of $D_{X}$, i.e., the inter-layer strength and, in the following, we focus on classical and diffusive random walks. 

In \cite{gomez2013diffusion} it has been shown that in the limit $D_{X}\lto\infty$ there are eigenvalues converging to a constant value and other eigenvalues diverging proportionally to $D_{X}$. The eigenvalues obtained from the normalized supra-Laplacian in the case of random walkers are related to the eigenvalues of the diffusion process by $\lambda_{\ell}\propto\lambda_{\ell}^{\text{Diff}}/D_{X}$. Substituting $\lambda_{2}\propto D_{X}^{-1}$ in \Eq{eq:rhovslambda2} we obtain that the time required to cover any given fraction of the multiplex is larger for increasing values of $D_{X}$. Our numerical experiments verify this theoretical expectation. An intuitive explanation is that when $D_{X}$ is much larger than the average vertex strength, the random walkers spend most of the time in switching layer instead of jumping to other vertices. In the specific case of RWP each switching action is followed by a jump within the same time step and, therefore, for this type of walk the time to cover a given fraction of the multiplex is not influenced by $D_{X}$. 

With a similar argument and the results obtained in \cite{gomez2013diffusion}, we have $\lambda_{2}\propto D_{X}$ when $D_{X}\lto 0$. This extremal case corresponds to a multiplex with vanishing inter-layer connections and the resulting coverage is no more dependent on the value of $D_{X}$, reducing \Eq{eq:rhovslambda2} to the coverage for a single layer.

\section{Dynamical vs Topological Resilience}

\begin{figure}[!h]
\centering
    \includegraphics[width=8.5cm]{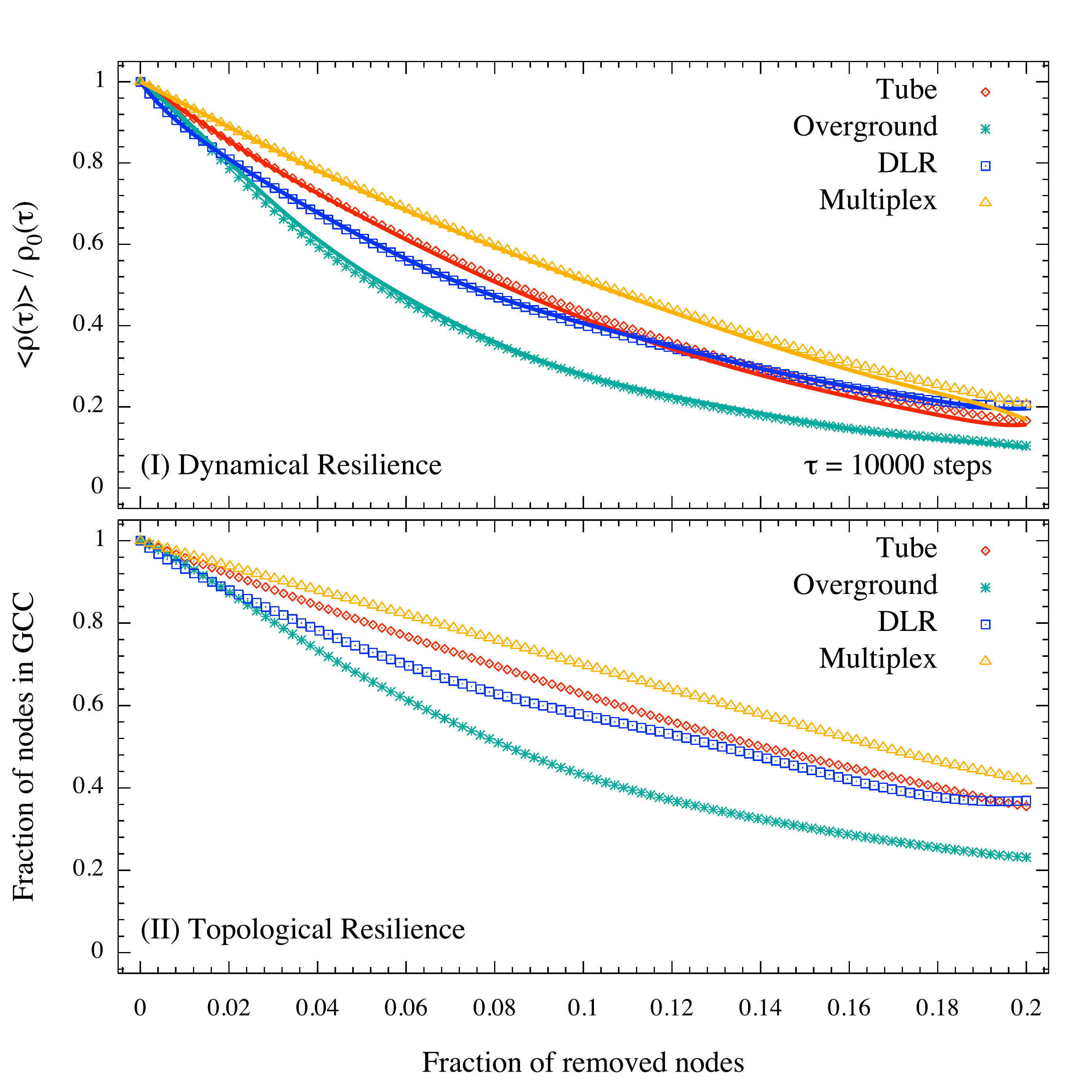}
 \caption{\label{fig:multiplex-london-resilience}\textbf{Resilience of the public transport network of London to random failures.\,} I) Theoretical expectations (solid lines) reproduce with great accuracy the resilience (points) obtained from simulations for each transportation layer and the whole interconnected system ($D_{X}=10^{-1}$), assuming random-walk based navigation. II) Structural resilience, defined as the average fraction of vertices surviving in the giant connected component after random failures.}
 \end{figure}

We capitalize on the presented theoretical framework to investigate the navigability resilience of interconnected networks to random failures, focusing on the particular case of the public transport of London. A failure, here, is considered as the inoperability of a station in a certain transportation layer (e.g. because of an accident, a traffic jam, or catastrophe). Such an event can happen randomly on the system and can affect one or more stations at the same time. A measure of the operability of the full system in response to unexpected failures, can be inferred from the coverage of the respective networks after such events. This is what we call the {\em navigability resilience}.
The resilience $r(\phi)$ of the system to a fraction $\phi$ or random failures is defined by $r(\phi)=\langle \rho_{\phi}(\tau) \rangle/\rho_{0}(\tau)$, where $\rho_{\phi}(t)$ is the coverage at time $\tau$ of the network subjected to $\phi$ failures and the averages are calculated over several random realizations of the failures. The normalization guarantees a fair comparison between the resilience of the multiplex and the monoplex networks. When a vertex fails in a single transportation layer, it can not be traversed by any path. However, if that vertex is part of an interconnected network it can be still reached on other layers. This intrinsic feature of multiplexes enhances the resilience of the system with respect to monoplexes, as shown in \Fig{fig:multiplex-london-resilience}-I for the public transport of London. 

We show in \Fig{fig:multiplex-london-resilience}-II the topological resilience corresponding to the same multiplex, defined by the average fraction of vertices surviving in the giant connected component after random failures. The navigability, i.e. the dynamical resilience, is inherently smaller than the topological resilience of this multiplex network. 

\newpage
\section{Empirical data of real disrupted services in London}

Finding information about possible disruptions in the transportation network of London, from the Oyster data in our possession (see the Main text for information), is very difficult. Moreover, it is difficult to collect information about disruptions occurring in 2009, the period in which our Oyster data refers to. For this reason, we have opted for collecting new data about disrupted services in London during a more recent period of time.

Our first choice has been the official data provided by Transport for London (TfL). Such data is provided in real time but, unfortunately, it concerns only ``Tube departure boards, line status and station status'', with no support for disruptions occurring to Overground and DLR, two out of three layers in the multiplex transportation network considered in this study. Moreover, it is not possible to access to historical disruptions.

For this reason, we decided to gather data from Twitter. In fact, delays and disruptions are reported in real time in this online social network by means of many different accounts, each one corresponding to a particular line. For our data collection, we considered tweets sent by the following accounts: TfLTravelAlerts, bakerlooline, metline, wlooandcityline, circleline, victorialine, LDNOverground, jubileeline, districtline, northernline, hamandcityline, LondonDLR and piccadillyline. We collected all the tweets containing the string ``no service'' in the message, sent from those accounts between 11 February 2012 and 26 March 2014. The two years of data guarantees a fair representation of the true distribution of disrupted services. Our choice is justified by the fact that we consider disrupted stations, not delays in the traffic. We collected more than 3000 tweets and, by means of conservative heuristics, we classified 64\% of them into 357 unique pairs of disrupted stations. 

Here, we report some representative examples of the latest tweets in our dataset, together with information about the account who sent the tweets and the date. Many tweets are just reply to other users:

\begin{verbatim}
Account:   LDNOverground   
Date:   15 mar 2014
Message:   
@alexandrafinlay there's no service on that line 
today. i advise you to de-select london overground
from the search. i'll pass this on too
\end{verbatim}

Such tweets are not used to classify disruptions. The rest of the tweets do not use a standard format and heuristics have been used to parse the information, conservatively. For instance, messages like

\begin{verbatim}
Account:   LDNOverground
Date:   9 mar 2014
Message:   no service between richmond - camden road, 
shepherds bush- willesden junction & watford junction- 
queens park due to planned  upgrade work.
\end{verbatim}
are difficult to be parsed, because the usage of symbols ``\&'' and ``-'' is somehow arbitrary. Nevertheless, our algorithm is able to recognize at least the disrupted pair ``richmond / camden road''. Apart from this type of tweets with ambiguous syntax, the majority of them have been correctly parsed. For instance, the algorithm correctly finds the multiple disrupted pairs ``euston / harrow\&wealdstone'', ``harrow\&wealdstone / watford'' in
\begin{verbatim}
Account:   LDNOverground
Date:   14 set 2014
Message:   (1 of 2) no service btn euston - harrow & 
wealdstone and severe delays btn harrow & 
wealdstone - watford junction.
\end{verbatim}
or ``bank / poplar'', ``bank / west india quay'', ``tower gateway / poplar'' and ``tower gateway / west india quay''
\begin{verbatim}
Account:   LondonDLR
Date:   21 set 2014
Message:   morning, ahmed & alex providing updates. 
due to planned work there is no service today between 
bank/tower gateway and poplar/west india quay
\end{verbatim}
or ``bank / canning town'', ``tower gateway / canning town'' and ``stratford / canary wharf'' in
\begin{verbatim}
Account:   LondonDLR
Date:   23 mar 2014
Message:   no service btn bank / tower gateway and 
canning town / canary wharf, and also between 
stratford and canary wharf. replacement buses operate.
\end{verbatim}
where ``btn'' and ``between'' are used for the same purpose. It is worth remarking here that this dataset is not intended to provide us with complete information about real disruptions occurring in London, but only to provide a fair sample of reasonable and most frequent disruptions, to be used as input in our simulations.  

The information about disruptions occurring to whole lines has been extracted manually from the data, without the usage of heuristics. However, for sake of completeness, we found reasonable to test all possible full-line disruptions (for a total of 11 possible disrupted multiplexes, excluding Overground and DLR which in our case are considered layers by themselves).

Here, we report details about some disruptions, ordered by their rank with respect to specific criteria. For instance, we consider:
\begin{itemize}
\item \textbf{Disruptions ranked by their frequency}. Here, frequency is calculated with respect to the data we have collected, and this is only a proxy for the true frequency of each disruption. Moreover, the most frequent disruptions are not, in general, the most dangerous for the traffic, involving only a limited amount of affected stations and often guaranteeing the connectedness of the underlying network. See Tab.\,\ref{tab:disruption-rank1}.
\item \textbf{Disruptions ranked by the number of stations they affect}. Here, disruptions might be more critical for the navigability of the system with respect to the previous ones. See Tab.\,\ref{tab:disruption-rank2}.
\item \textbf{Whole-line disruptions}. Disruption of a complete tube line is considered in each scenario, for a total of 11 lines. See Tab.\,\ref{tab:line-disruption-results}.
\end{itemize}

In Tab.\,\ref{tab:disruption-results} we report the dynamical resilience calculated, numerically and theoretically, for some representative real partial disruptions, mainly sampled from Tab.\,\ref{tab:disruption-rank1} and Tab.\,\ref{tab:disruption-rank2}.  In Tab.\,\ref{tab:line-disruption-results} we report the same analysis for disruptions of whole lines. The values of the data-driven simulations are in remarkable agreement with our theory.

% \newpage
\begin{table*}[!t]
\caption{\label{tab:disruption-rank1} Real disruptions in London transportation network, ranked by their occurrence in our dataset. The partially disrupted line (``Line'' column) is reported, together with the starting (``From'' column) and ending (``To'' column) stations affected by the disruption. The rate of occurrence (``Freq.'' column) is also reported together with the fraction of stations indirectly affected (``Affected'' column).}
\scalebox{1}{
\begin{tabular}{llllllllll}
\hline\hline
\textbf{ID} & \textbf{Line} & \textbf{From} & \textbf{To} & \textbf{Freq.} & \textbf{Affected} \\\hline\hline
DISR1 & metropolitan & aldgate & bakerstreet & 3.35\% & 2.44\%\\\hline
DISR4 & overground & claphamjunction & surreyquays & 2.02\% & 1.90\%\\\hline
DISR3 & dlr & beckton & canningtown & 2.56\% & 2.44\%\\\hline
DISR2 & hammersmith\&city & barking & moorgate & 2.89\% & 3.52\%\\\hline
DISR9 & piccadilly & raynerslane & uxbridge & 1.53\% & 1.90\%\\\hline
DISR8 & overground & claphamjunction & willesdenjunction & 1.57\% & 1.63\%\\\hline
DISR7 & piccadilly & actontown & uxbridge & 1.57\% & 4.07\%\\\hline
DISR6 & northern & edgware & hampstead & 1.82\% & 1.90\%\\\hline
DISR5 & overground & richmond & willesdenjunction & 1.94\% & 1.63\%\\\hline
DISR26 & metropolitan & aldgate & wembleypark & 1.07\% & 2.98\%\\\hline
DISR25 & overground & richmond & stratford & 1.07\% & 6.23\%\\\hline
DISR24 & metropolitan & aldgate & harrow-on-the-hill & 1.12\% & 3.79\%\\\hline
DISR23 & district & ealingbroadway & turnhamgreen & 1.11\% & 1.36\%\\\hline
DISR22 & overground & highbury\&islington & newcross & 1.11\% & 3.52\%\\\hline
DISR21 & overground & camdenroad & richmond & 1.16\% & 4.07\%\\\hline
DISR20 & northern & camdentown & kennington & 1.16\% & 2.71\%\\\hline
DISR19 & metropolitan & northwood & wembleypark & 1.16\% & 2.17\%\\\hline
DISR18 & dlr & bowchurch & stratford & 1.20\% & 0.81\%\\\hline
DISR17 & overground & sydenham & westcroydon & 1.24\% & 1.36\%\\\hline
DISR16 & northern & camdentown & millhilleast & 1.28\% & 2.17\%\\\hline
\end{tabular}
}
\end{table*}

\newpage
\begin{table*}[!t]
\caption{\label{tab:disruption-rank2} Real disruptions in London transportation network, ranked by the number of stations they affect. The partially disrupted line (``Line'' column) is reported, together with the starting (``From'' column) and ending (``To'' column) stations affected by the disruption. The rate of occurrence (``Freq.'' column) is also reported together with the fraction of stations indirectly affected (``Affected'' column).}
\scalebox{1}{
\begin{tabular}{llllllllll}
\hline\hline
\textbf{ID} & \textbf{Line} & \textbf{From} & \textbf{To} & \textbf{Freq.} & \textbf{Affected} \\\hline\hline
DISR325 & northern & eastfinchley & morden & 0.04\% & 7.05\%\\\hline
DISR281 & northern & goldersgreen & morden & 0.04\% & 6.78\%\\\hline
DISR25 & overground & richmond & stratford & 1.07\% & 6.23\%\\\hline
DISR245 & piccadilly & actontown & arnosgrove & 0.04\% & 6.78\%\\\hline
DISR88 & northern & edgware & kennington & 0.33\% & 5.15\%\\\hline
DISR61 & overground & claphamjunction & stratford & 0.45\% & 5.96\%\\\hline
DISR44 & overground & highbury\&islington & westcroydon & 0.58\% & 5.69\%\\\hline
DISR347 & district & earlscourt & westham & 0.04\% & 5.69\%\\\hline
DISR322 & overground & southacton & stratford & 0.04\% & 5.42\%\\\hline
DISR250 & jubilee & stratford & willesdengreen & 0.04\% & 5.42\%\\\hline
DISR227 & metropolitan & aldgate & rickmansworth & 0.08\% & 5.42\%\\\hline
DISR220 & overground & hackneywick & richmond & 0.08\% & 5.96\%\\\hline
DISR199 & metropolitan & aldgate & croxley & 0.08\% & 5.42\%\\\hline
DISR195 & district & towerhill & upminster & 0.08\% & 5.42\%\\\hline
DISR184 & northern & millhilleast & stockwell & 0.08\% & 5.15\%\\\hline
DISR181 & northern & highbarnet & stockwell & 0.08\% & 5.96\%\\\hline
DISR180 & metropolitan & aldgate & uxbridge & 0.08\% & 5.96\%\\\hline
DISR175 & hammersmith\&city & bakerstreet & barking & 0.12\% & 5.15\%\\\hline
DISR151 & district & embankment & upney & 0.12\% & 5.15\%\\\hline
DISR140 & northern & highbarnet & kennington & 0.17\% & 5.42\%\\\hline
\end{tabular}
}
\end{table*}

\newpage
\begin{table*}[!t]
\caption{\label{tab:disruption-results} Real partial disruptions in the London transportation network. Representative disruptions are considered, together with the starting (``From'' column) and ending (``To'' column) stations affected by the disruption. The rate of occurrence (``Freq.'' column) is reported, together with the fraction of stations indirectly affected (``Affected'' column). It is indicated if the resulting multiplex is disconnected in 2 or more components (``Discon.?'' column). The resilience obtained from Monte Carlo simulations (random walk and shortest-path based) are reported together with our theoretical expectation.}
\scalebox{0.84}{
\begin{tabular}{llllllclll}
\hline\hline
\textbf{ID} & \textbf{Line} & \textbf{From} & \textbf{To} & \textbf{Freq.} & \textbf{Affected} & \textbf{Discon.?} & \textbf{Th.Res.} & \textbf{RW Res.} & \textbf{SP Res.}\\\hline\hline
DISR1 & metropolitan & aldgate & bakerstreet & 3.35\% & 2.44\% & NO & 99.60\% & 100\% & 99.99\% \\\hline
DISR4 & overground & claphamjunction & surreyquays & 2.02\% & 1.90\% & YES & 92.34\% & 90.56\% &100\%\\\hline
DISR3 & dlr & beckton & canningtown & 2.56\% & 2.44\% & YES & 94.41\% & 93.10\% & 94.85\%\\\hline
DISR325 & northern & eastfinchley & morden & 0.04\% & 7.05\% & YES & 85.90\% & 82.52\% & 87.07\%\\\hline
DISR25 & overground & richmond & stratford & 1.07\% & 6.23\% & YES & 89.07\% & 91.53\% & 97.90\%\\\hline
DISR245 & piccadilly & actontown & arnosgrove & 0.041\% & 6.78\% & YES & 88.50\% & 86.51\% & 90.85\%\\\hline
DISR181 & northern & highbarnet & stockwell & 0.083\% & 5.96\% & YES & 86.11\% & 82.55\% & 84.49\%\\\hline
DISR61 & overground & claphamjunction & stratford & 0.45\% & 5.96\% & YES & 86.59\% & 84.99\% & 99.66\%\\\hline
DISR119 & northern & charingcross & highbarnet & 0.25\% & 4.61\% & YES & 90.67\% & 87.99\% & 91.32\%\\\hline
\end{tabular}
}
\end{table*}
%\clearpage
\newpage

\begin{table*}[!t]
\caption{\label{tab:line-disruption-results} Complete line disruptions in London transportation network. Same as table 4, but also indicating if the resulting multiplex is disconnected in 2 or more components (``Discon.?'' column). The resilience obtained from Monte Carlo simulations (random walk and shortest-path based) are reported together with our theoretical expectation.}
\scalebox{0.84}{
\begin{tabular}{lllclll}
\hline\hline
\textbf{ID} & \textbf{Line} & \textbf{Affected} & \textbf{Discon.?} & \textbf{Th.Res.} & \textbf{RW Res.} & \textbf{SP Res.}\\\hline\hline
DISR-L1 & bakerloo & 6.78\% & YES & 95.80\% & 96.25\% & 99.79\%\\\hline
DISR-L2 & circle & 9.49\% & NO & 99.68\% & 100\% & 99.93\%\\\hline
DISR-L3 & district & 16.26\% & YES & 89.37\% & 89.47\% & 96.61\%\\\hline
DISR-L4 & hammersmith\&city & 7.86\% & YES & 99.18\% & 99.460\% & 99.71\%\\\hline
DISR-L5 & jubilee & 7.32\% & YES & 91.50\% & 93.08\% & 100\%\\\hline
DISR-L6 & metropolitan & 9.21\% & YES & 91.96\% & 91.53\% & 95.43\%\\\hline
DISR-L7 & northern & 13.55\% & YES & 84.41\% & 80.98\% & 89.51\%\\\hline
DISR-L8 & piccadilly & 14.36\% & YES & 85.07\% & 83.43\% & 91.23\%\\\hline
DISR-L9 & victoria & 4.34\% & YES & 95.33\% & 96.78\% & 100\%\\\hline
DISR-L10 & central & 13.27\% & YES & 83.35\% & 80.49\% & 90.00\%\\\hline
DISR-L11 & waterloo\&city & 0.54\% & NO & 99.98\% & 100\% & 100\%\\\hline
\end{tabular}
}
\end{table*}

\clearpage
\bibliographystyle{naturemag}
\bibliography{navigability_multiplex}

\begin{thebibliography}{10}
\expandafter\ifx\csname url\endcsname\relax
  \def\url#1{\texttt{#1}}\fi
\expandafter\ifx\csname urlprefix\endcsname\relax\def\urlprefix{URL }\fi
\providecommand{\bibinfo}[2]{#2}
\providecommand{\eprint}[2][]{\url{#2}}

\bibitem{stanley83}
\bibinfo{author}{Stanley, H.~E.}, \bibinfo{author}{Kang, K.},
  \bibinfo{author}{Redner, S.} \& \bibinfo{author}{Blumberg, R.~L.}
\newblock \bibinfo{title}{Novel superuniversal behavior of a random-walk
  model}.
\newblock \emph{\bibinfo{journal}{Phys. Rev. Lett.}}
  \textbf{\bibinfo{volume}{51}}, \bibinfo{pages}{1223--1226}
  (\bibinfo{year}{1983}).

\bibitem{sinatra2011maximal}
\bibinfo{author}{Sinatra, R.}, \bibinfo{author}{G{\'o}mez-Garde{\~n}es, J.},
  \bibinfo{author}{Lambiotte, R.}, \bibinfo{author}{Nicosia, V.} \&
  \bibinfo{author}{Latora, V.}
\newblock \bibinfo{title}{Maximal-entropy random walks in complex networks with
  limited information}.
\newblock \emph{\bibinfo{journal}{Phys. Rev. E}} \textbf{\bibinfo{volume}{83}},
  \bibinfo{pages}{030103} (\bibinfo{year}{2011}).

\bibitem{wilson1972introduction}
\bibinfo{author}{Wilson, R.~J.}
\newblock \emph{\bibinfo{title}{Introduction to graph theory}}, vol.
  \bibinfo{volume}{111} (\bibinfo{publisher}{Academic Press New York},
  \bibinfo{year}{1972}).

\bibitem{tetali1991random}
\bibinfo{author}{Tetali, P.}
\newblock \bibinfo{title}{Random walks and the effective resistance of
  networks}.
\newblock \emph{\bibinfo{journal}{J. Theor. Probab.}}
  \textbf{\bibinfo{volume}{4}}, \bibinfo{pages}{101--109}
  (\bibinfo{year}{1991}).

\bibitem{noh2004random}
\bibinfo{author}{Noh, J.~D.} \& \bibinfo{author}{Rieger, H.}
\newblock \bibinfo{title}{Random walks on complex networks}.
\newblock \emph{\bibinfo{journal}{Phys. Rev. Lett.}}
  \textbf{\bibinfo{volume}{92}}, \bibinfo{pages}{118701}
  (\bibinfo{year}{2004}).

\bibitem{yang2005exploring}
\bibinfo{author}{Yang, S.-J.}
\newblock \bibinfo{title}{Exploring complex networks by walking on them}.
\newblock \emph{\bibinfo{journal}{Phys. Rev. E}} \textbf{\bibinfo{volume}{71}},
  \bibinfo{pages}{016107} (\bibinfo{year}{2005}).

\bibitem{samukhin2008laplacian}
\bibinfo{author}{Samukhin, A.}, \bibinfo{author}{Dorogovtsev, S.} \&
  \bibinfo{author}{Mendes, J.}
\newblock \bibinfo{title}{Laplacian spectra of, and random walks on, complex
  networks: Are scale-free architectures really important?}
\newblock \emph{\bibinfo{journal}{Phys. Rev. E}} \textbf{\bibinfo{volume}{77}},
  \bibinfo{pages}{036115} (\bibinfo{year}{2008}).

\bibitem{burda2009localization}
\bibinfo{author}{Burda, Z.}, \bibinfo{author}{Duda, J.}, \bibinfo{author}{Luck,
  J.} \& \bibinfo{author}{Waclaw, B.}
\newblock \bibinfo{title}{Localization of the maximal entropy random walk}.
\newblock \emph{\bibinfo{journal}{Phys. Rev. Lett.}}
  \textbf{\bibinfo{volume}{102}}, \bibinfo{pages}{160602}
  (\bibinfo{year}{2009}).

\bibitem{watts1998collective}
\bibinfo{author}{Watts, D.} \& \bibinfo{author}{Strogatz, S.}
\newblock \bibinfo{title}{Collective dynamics of ``small-world'' networks}.
\newblock \emph{\bibinfo{journal}{Nature}} \textbf{\bibinfo{volume}{393}},
  \bibinfo{pages}{440--442} (\bibinfo{year}{1998}).

\bibitem{gomez2013diffusion}
\bibinfo{author}{G{\'o}mez, S.} \emph{et~al.}
\newblock \bibinfo{title}{Diffusion dynamics on multiplex networks}.
\newblock \emph{\bibinfo{journal}{Phys. Rev. Lett.}}
  \textbf{\bibinfo{volume}{110}}, \bibinfo{pages}{028701}
  (\bibinfo{year}{2013}).

\end{thebibliography}

%\end{small}
%\end{multicols}
\end{document}